\documentclass[12pt,letterpaper]{JHEP3}
\usepackage{amsmath,amssymb,amsfonts,xspace}
\usepackage{epsfig}

\newcommand{\color}[6]{}  

\numberwithin{equation}{section}
\def\det{\,{\rm det}\, }

\def\Im{\,{\rm Im}\,}
\def\Re{\,{\rm Re}\,}

\def\({\left(}
\def\){\right)}
\def\[{\left[}
\def\]{\right]}
\def\haf{\textstyle{1\over 2}}
\def\hf{\frac{1}{2}}

\renewcommand{\d}{\mathrm{d}}
\newcommand{\de}{\mathrm{d}}

\newcommand{\I}{\mathrm{i}}

\newcommand{\cL}{\mathcal{L}}

\newcommand{\p}{\partial}

\newcommand{\half}{\frac{1}{2}}

\newcommand{\cF}{\mathcal{F}}
\newcommand{\cV}{\mathcal{V}}

\newcommand{\cC}{\mathcal{C}}
\newcommand{\cS}{\mathcal{S}}

\newcommand{\cM}{\mathcal{M}}

\newcommand{\cX}{\mathcal{X}}
\newcommand{\cY}{\mathcal{Y}}
\newcommand{\CX}{\mathcal{X}}
\newcommand{\cR}{\mathcal{R}}

\DeclareSymbolFont{AMSa}{U}{msa}{m}{n}
\DeclareSymbolFont{AMSb}{U}{msb}{m}{n}
\DeclareMathSymbol{\fieldR}{\mathalpha}{AMSb}{"52}

\newcommand{\kahler}{{K\"ahler}\xspace}
\newcommand{\hk}{{hyperk\"ahler}\xspace}
\newcommand{\qk}{{quaternion-K\"ahler}\xspace}

\newcommand{\cZ}{\mathcal{Z}}

\newcommand{\cO}{\mathcal{O}}

\newcommand{\cU}{\mathcal{U}}
\newcommand{\cA}{\mathcal{A}}

\newcommand{\pa}{\partial}
\newcommand{\nn}{\nonumber}

\newcommand{\eps}{\epsilon}
\newcommand{\IR}{\mathbb{R}}
\newcommand{\IC}{\mathbb{C}}
\newcommand{\IZ}{\mathbb{Z}}
\newcommand{\IN}{\mathbb{N}}
\newcommand{\IH}{\mathbb{H}}

\newcommand{\tzeta}{{\tilde\zeta}}

\newcommand{\txi}{{\tilde\xi}}

\newcommand{\CP}{\IC P^1}

\def\bea{\begin{eqnarray}}
\def\eea{\end{eqnarray}}
\def\be{\begin{equation}}
\def\ee{\end{equation}}
\def\ba{\begin{align}}
\def\ea{\end{align}}
\def\bse{\begin{subequations}}
\def\ese{\end{subequations}}

\def\ba{\bar a}

\def\bz{\bar z}

\def\Hp{H_{\scriptscriptstyle{\smash{(1)}}}}

\def\ui#1{^{[#1]}}
\def\di#1{_{[#1]}}

\def\txii#1{{\tilde\xi}^{[#1]}}
\def\ai#1{{\alpha}^{[#1]}}

\def\xii#1{\xi^{[#1]}}

\def\Hij#1{H^{[#1]}}

\def\Hpij#1{\Hij{#1}_{\scriptscriptstyle{\smash{(1)}}}}

\def\muh{\mu}
\def\hSij#1{S^{[#1]}}
\def\hHij#1{H^{[#1]}}

\newcommand{\hCX}{\mathcal{X}}

\def\cij#1{c}
\def\ci#1{c}

\def\balp{\bar \alpha}

\def\bxi{\bar \xi}
\def\btxi{\bar {\tilde \xi}}

\def\ba{\bar a}

\def\bt{\bar t}

\def\bw{\bar w}

\def\bz{\bar z}

\def\tc{c_{\txi}}
\def\ac{c_\alpha}
\def\varpi{t}

\def\tT{{\tilde T}}

\def\Prz{{\rm Prz}}
\def\dPrz{{\rm dPrz}}
\def\todaQ{T}
\def\cpF{\mathcal{P}}
\def\cpphi{\varphi}
\def\cprho{{\tau}}
\def\varsigmapar{\nu}

\title{Self-dual Einstein Spaces, Heavenly Metrics and Twistors}

\preprint{LPTA/09-099, ITP-UU-09-58,\\SPIN-09-48}

\author{Sergei Alexandrov$^1$, Boris Pioline$^{2}$,
Stefan Vandoren$^{2,3}$
\\
$^1$ {\it Laboratoire de Physique Th\'eorique \&
Astroparticules, CNRS UMR 5207, \\
Universit\'e Montpellier II, 34095 Montpellier Cedex 05, France}\\

$^2$ {\it Laboratoire de Physique Th\'eorique et Hautes
Energies, CNRS UMR 7589, \\
Universit\'e Pierre et Marie Curie,
4 place Jussieu, 75252 Paris cedex 05, France} \\

$^3$ {\it   Institute for Theoretical Physics and
           Spinoza Institute,
           Utrecht University,
           Leuvenlaan 4,
           3508 TD Utrecht,
           The Netherlands
           }

\vspace*{2mm} {\tt e-mail: \email{alexandrov@lpta.univ-montp2.fr},
\email{pioline@lpthe.jussieu.fr},
\email{S.J.G.Vandoren@uu.nl}} \vspace*{-3mm}

}

\abstract{
Four-dimensional quaternion-K\"ahler metrics, or equivalently self-dual Einstein
spaces $\cM$, are known to be encoded locally into one real function $h$ subject to
Przanowski's Heavenly equation. We elucidate the relation between
this description and the usual twistor description
for quaternion-K\"ahler spaces. In particular, we show that the same space $\cM$
can be described by infinitely many different solutions $h$, associated to different
complex (local) submanifolds on the twistor space, and therefore to different
(local) integrable complex structures on $\cM$.  We also study quaternion-K\"ahler
deformations of $\cM$ and, in the special case where $\cM$ has a Killing vector field,
show that the corresponding variations of $h$
are related to eigenmodes of the conformal Laplacian on $\cM$.
We exemplify our findings on the four-sphere $S^4$, the hyperbolic plane $H^4$
and on the ``universal hypermultiplet", i.e. the hypermultiplet moduli space in type IIA string
compactified on a rigid Calabi-Yau threefold.}

\begin{document}

\section{Introduction and summary}

Einstein spaces with self-dual Weyl curvature have been much studied both in the mathematics
and physics literature. On the mathematics side, they embody the notion
of quaternion-K\"ahler geometry in four dimensions, and are thus amenable to the same
algebro-geometric, twistorial techniques as their higher-dimensional cousins \cite{besse1987em}.
In the case of compact Einstein spaces  with positive scalar curvature, the self-duality constraint is so strong
as to leave only two possibilities: the four-sphere $S^4$ equipped with the round metric,
and  the complex projective plane $\IC P^2=SU(3)/U(2)$ equipped with the Fubini-Study metric. For
non-positive curvature however (the zero-curvature case corresponding to hyperk\"ahler manifolds),
the situation is much richer and a classification is still lacking.

On the physics side, self-dual Einstein spaces are the natural notion
of gravitational instanton in the presence of a cosmological constant. In the case of negative
curvature, these spaces can also be used as target spaces for locally supersymmetric
sigma models with 8 supercharges \cite{Bagger:1983tt}. A prominent example is the hypermultiplet moduli
space $\cM_H(X)$ for type IIA string  theory compactified on a rigid Calabi-Yau manifold $X$, which has
received much attention in recent years \cite{Strominger:1997eb,Becker:1999pb,Gutperle:2000sb,
Ketov:2002vr,Antoniadis:2003sw,Anguelova:2004sj,Davidse:2004gg,Davidse:2005ef,Alexandrov:2006hx,Bao:2009fg}.
Computing the exact self-dual Einstein metric on $\cM_H(X)$ including all instanton corrections
is still an open problem, which motivates the present study.

As shown long ago by Przanowski \cite{Przanowski:1984qq}, self-dual Einstein metrics $\cM$
can be described locally by one real function $h$ on an open neighborhood $U$ in  $\cM$
coordinatized by complex coordinates $z^\alpha$, $\alpha=1,2$, subject to the partial differential equation
\be
\Prz(h)\equiv h_{1\bar 1} h_{2\bar 2}
-h_{1\bar 2}h_{{\bar 1} 2}+\(2h_{1\bar 1}-h_1 h_{\bar 1}\){\rm e}^h=0\, ,
\label{master}
\ee
where, $h_1\equiv \partial h/\partial z^1$, etc. Any solution of this  equation then
leads to a self-dual Einstein metric on $\cU$ given by
\be
{\rm d}s^2_\cM= -\frac{6}{\Lambda}\(h_{\alpha \bar\beta}  \, {\rm d}z^{\alpha}\, {\rm d}z^{\bar\beta} +
2  {\rm e}^h \, | \de z^2|^2  \)
\equiv 2 g_{\alpha\bar\beta}
\,{\rm d}z^{\alpha}\, {\rm d}z^{\bar\beta}\, ,
\label{metPrz}
\ee
normalized such that the Ricci scalar satisfies the standard convention in four dimensions
$R=4\Lambda$. The ``master equation"
\eqref{master} is a variant of Plebanski's ``first heavenly equation", which
similarly parametrizes \hk metrics in terms of one real function on an open
set in $\IC^2$. It  is crucial to note that  \eqref{metPrz} exhibits the
metric in Hermitian form with respect to a local (integrable) complex structure $J_h$ with complex
coordinates $z^1,z^2$. This in general can only hold locally: indeed,  self-dual Einstein
spaces generally have no globally defined complex structures, as exemplified by $S^4$.

The Przanowski form of the metric \eqref{metPrz} provides a concise
way of summarizing the constraints of \qk geometry in 4 dimensions. Moreover,  it offers
a convenient starting point for analyzing perturbations  of $\cM$ consistent with the self-dual
Einstein property \cite{Alexandrov:2006hx}: indeed, such perturbations correspond to
solutions $\delta h$ of the ``linearized master equation"
\be
\dPrz_h (\delta h)=0\, ,
\label{eqdPrz}
\ee
where $h$ is a solution of \eqref{master}, $\delta h$ its infinitesimal variation,
and we defined the linearized Przanowski operator
\be
\label{dPrz}
\dPrz_h\equiv
(h_{2\bar 2}+2\, {\rm e}^h)\p_1\p_{\bar 1}+h_{1\bar 1}\p_2\p_{\bar 2}-
h_{1\bar 2}\p_2\p_{\bar 1}-h_{2\bar 1}\p_1\p_{\bar 2}
+{\rm e}^h(2h_{1\bar 1}-|h_1|^2-h_1\p_{\bar 1}-h_{\bar 1}\p_1)\, .
\ee
Of course, there may be
obstructions to integrating a solution of \eqref{dPrz}  to a smooth deformation of $\cM$, as must be the
case for compact positively curved manifolds. Moreover, as will be apparent presently,
some of the solutions of \eqref{dPrz} may be ``pure gauge", i.e. may describe the same metric $\cM$
up to  diffeomorphism.

Indeed, there are in general infinitely many ways  of expressing a given self-dual
Einstein metric in Przanowski form \eqref{metPrz}. One obvious source of infinities,
comes from holomorphic changes of variables of the restricted form \cite{Przanowski:1990gf},
\begin{equation}
\label{diff-equiv}
z^1\rightarrow {z'}^{1} = f(z^1,z^2)\, ,
\qquad
z^2 \rightarrow {z'}^{2} = g(z^2)\, ,
\end{equation}
\begin{equation}
\label{equiv-h}
h(z^1,z^2) \rightarrow {h'}(z^1,z^2) =
h\(f(z^1,z^2),g(z^2)\) - \log \left( g_2(z^2)\,
{\bar g}_{\bar 2}(\bar {z}^2) \right)\, .
\end{equation}
This change of variables
preserves the local complex structure $J_h$ and leads to the same metric up to diffeomorphism.
At the infinitesimal level, the holomorphic change of variables \eqref{diff-equiv} leads to ``pure
gauge" solutions  of the linearized equation \eqref{dPrz}, of the form
\be
\label{pgt}
\dPrz_h\left(  h_1\,\delta f + h_2 \,\delta g+ \delta g_2  \right)
= (\delta f_1 + \delta f \, \p_1+ \delta g_2 + \delta g\, \p_2)\,  \Prz(h)=0\, ,
\ee
where $\delta f(z^1, z^2)$ and $\delta g(z^2)$ are infinitesimal versions of \eqref{diff-equiv}.

More significantly, as emphasized by Tod \cite{Tod:2006wj}
and reviewed later in this work,
the Przanowski Ansatz \eqref{metPrz} can be reached for any choice of local integrable complex
structure $J$. In particular, different local integrable complex structures $J$ and $J'$
lead to different solutions $h(z^1,z^2)$ and $h'(z'^1,z'^2)$ of the master equation, such that
$J=J_{h}$ and $J'=J_{h'}$. These solutions describe the same self-dual Einstein metric,
but the coordinate systems $(z^1,z^2)$ and $(z'^1,z'^2)$ are in general related by a non-holomorphic
change of variables, as depicted in Figure 1 (to be elaborated on below).
Quaternion-\kahler spaces admit infinitely many integrable local complex structures,
which provides infinitely many ways of expressing the same self-dual Einstein metric in
Przanowski's form. Unfortunately, it is not known how to express the corresponding ``pure gauge"
solutions $\delta h$ of \eqref{eqdPrz} in terms of the unperturbed solution $h$ and its derivatives
as in \eqref{pgt}.

Our purpose in this note is to make contact between the Przanowski parametrization of
self-dual Einstein spaces, reviewed above, and the standard twistor description of \qk spaces,
recalled below. In doing so, we shall address the problem just raised, and provide a way to generate
all solutions of the master equation which describe the same self-dual Einstein space using twistor techniques.

\begin{figure}
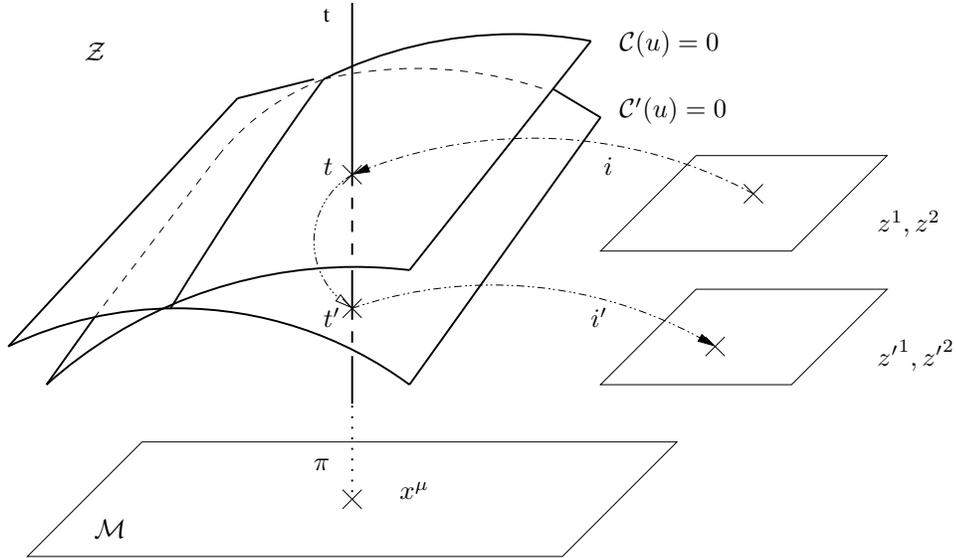

\centerline{\input prza.pspdftex}
\caption{Two Przanowski coordinate systems $(z^1, z^2)$, $(z'^1, z'^2)$ associated to
two different submanifolds $\cC,\cC' \subset \cZ$ are related by a non-holomorphic coordinate
change. The relation is found by intersecting the twistor line through
$i(z^1, z^2)\in \cC$ (the vertical line in this figure) with $\cC'$.\label{fig1}
}
\end{figure}

\subsection{Preview of main results}

To preview our main result, recall that for any $4n$-dimensional \qk space $\cM$,
one may construct its twistor space $\cZ$, a two-sphere bundle over $\cM$
which admits a canonical integrable complex structure together with a complex contact structure.
Any local section of this bundle gives rise to an almost complex structure on an open subset $U=\pi(\cU)\subset \cM$.
The key fact for us is that this (local)
almost complex structure is integrable if and only if the submanifold $\cC$ defined by the section is a (local)
complex submanifold on $\cU\subset \cZ$ \cite{quatman}.

On the other hand, the \qk
metric on $\cM$ is related to the \kahler-Einstein metric on $\cZ$ via
\begin{equation}
\label{dsMZ}
\frac{\Lambda}{12}\,{\rm d}s^2_{\cM}={\rm d}s^2_{\cZ}-{\rm e}^{-2K}|\cX|^2\, ,
\end{equation}
where $\cX$ is the (local) complex contact one-form and $K$ is the (local) \kahler potential
on $\cZ$. Restricting \eqref{dsMZ} to the submanifold $\cC$ and identifying $h_\cC=-2 K$ on that
slice, we can express the \qk metric on $\cM$ in the Hermitian form \cite{deWit:2001dj}
\be
{\rm d}s^2_{\cM}= -\frac{6}{\Lambda}\(\p_{\alpha}\p_{\bar\beta} h_\cC \, {\rm d}z^{\alpha}\, {\rm d}{\bar z}^{\bar\beta} +
2  {\rm e}^{h_\cC}  | \cX_{\cC}|^2  \)\, ,
\label{metPrzg}
\ee
where $z^\alpha, \alpha=1,\dots ,2n$ are complex coordinates along $\cC$ and $\cX_{\cC}$
is the restriction of the contact one-form to $\cC$.  For $n=1$, one may always choose complex
coordinates $z^1,z^2$ and perform a suitable \kahler transformation such that $\cX\vert_{\cC}=\de z^2$,
hence recovering the Przanowski Ansatz \eqref{metPrz}. As for the master equation \eqref{master},
it can be shown to follow from the Monge-Amp\`ere equation\footnote{Observe that, except for the last
step, the argument just outlined holds in arbitrary dimension: for $n> 1$,
it is still possible to parametrize
the most general $4n$-dimensional \qk manifold in Przanowski-type form \eqref{metPrzg}, with
$\cX_\cC$ chosen in one's favorite Darboux form such as $\cX_\cC=\sum_{i=1}^{n} z^{2i-1} \de z^{2i}$.
However, the constraints of \qk geometry will involve
$2n-1$ partial differential equations rather
than a single one.} on $\cZ$. We have therefore identified the twistorial origin of
the Przanowski function $h_\cC$: it is just a ``tomogram" of the \kahler potential
$K$ along the complex submanifold $\cC$. Varying the choice of locally integrable
complex structure $J$ on $\cM$, therefore the complex submanifold $\cC$ of
$\cZ$, one can in principle reconstruct the \kahler potential on the whole of  $\cZ$.

Having established the twistorial origin of the Przanowski function, we can
cast the solutions of the linearized master equation \eqref{eqdPrz}
 into the formalism developed in our previous work \cite{Alexandrov:2008nk}
 for linear deformations of \qk manifolds.
To this end, we use the fact from \cite{Alexandrov:2008nk}  that deformations of the \kahler potential
$K$ on $\cZ$ (or rather, of the contact potential, defined in \eqref{c-pot} and
\eqref{Kpotential} below) can be written as a contour integral of a certain holomorphic
function (more precisely a section $\Hp$ of the \v{C}ech cohomology group $H^1(\cZ,\cO(2))$),
which encodes the variation of the contact transformations between
various local Darboux coordinate systems on $\cZ$. In the special case where the
unperturbed metric $\cM$ has a Killing vector, this leads to a contour integral
formula \eqref{delh} for the corresponding deformation $\delta h$ of the Przanowski function.
This formula is very similar to the Penrose-type formula \eqref{Pen1} which
produces eigenmodes of the
conformal Laplacian from sections $\Psi$ in $H^1(\cZ,\cO(-2))$ (as discussed in \cite{Neitzke:2007ke}).
This is not an accident because, when $\cM$
has a Killing vector, the linearized master equation \eqref{eqdPrz} turns out
to be equivalent to the conformal Laplace equation for the rescaled function
$\delta h/ |h_1|^2$, as explained in \eqref{geneqiso} below. In this fashion, we can
identify the holomorphic sections $\Hp$ and $\Psi$ in the particular trivialization
introduced in  \cite{Neitzke:2007ke}.

The rest of this note aims at substantiating the above claims and illustrating them on various examples.
It is organized as follows: In Section \ref{sec_review}, we review the twistorial description of \qk spaces.
In Section \ref{sec_PrfromT}, we explain how all solutions of the master equation can be obtained by restricting
the \kahler potential on the twistor space on any local complex submanifold, provide
explicit formulae for relating different classes of solutions, and discuss how to recover the full twistor space
from a single ``tomogram".
In Section \ref{sec_isometr} we consider self-dual Einstein spaces with one or two
commuting isometries and relate the respective Toda and Calderbank-Pedersen descriptions
to the Przanowski framework.
In Section \ref{sec_examples} we illustrate these results by revisiting the $S^4$ and
$H^4$ examples, as well as a deformed, non-compact version of $\IC P^2$
which describes the perturbative hypermultiplet moduli space of Type IIA string theory
compactified on a rigid Calabi-Yau threefold.
Finally, in Section \ref{seclin}, we present one of our main results,
a Penrose-type formula which produces a solution to the linearized master equation
\eqref{eqdPrz} from an element of $H^1(\cZ,\cO(-2))$ (locally, a holomorphic function on $\cZ$),
and explain the relation with the $H^1(\cZ,\cO(2))$ section which governs the perturbations
of $\cZ$. These results are illustrated on instanton corrections to the
afore-mentioned hypermultiplet moduli space.

\section{Review of the twistor approach to quaternionic geometries}
\label{sec_review}

In real dimension $4n\geq 8$, a quaternion-K\"ahler manifold $\cM$ is a
Riemannian manifold with  metric $g_\cM$ and Levi-Civita connection
$\nabla$ such that the holonomy group is contained in $USp(n)\times SU(2)$, see e.g.
\cite{MR664330}. In dimension $4n=4$, the case of interest in this paper,
this holonomy condition is trivially satisfied, and must be replaced by the
condition that the metric be Einstein and the Weyl curvature be
self-dual. In any dimension, quaternion-K\"ahler manifolds can also be defined
as Riemannian manifolds that admit a triplet of almost complex Hermitian
structures $\vec J$ (defined up to $SU(2)$ rotations) satisfying the algebra of the unit imaginary
quaternions. The quaternionic two-forms
$\vec\omega_\cM(X,Y)\equiv g_\cM(\vec J X,Y)$ must be covariantly closed with
respect to the $SU(2)$ part  $\vec p$ of the Levi-Civita connection
and be proportional to the curvature
of $\vec p$,
\be \label{ompp}
{\rm d}\vec\omega_\cM + \vec p \times
\vec\omega_\cM = 0\, ,
\qquad
{\rm d}\vec p+ \half \,\vec p \times \vec
p = \frac{\nu}{2}\, \vec \omega_\cM\, ,
\ee
where we use the notation
$(\vec v \times \vec w)^a=\epsilon^{abc}\,v^b\wedge w^c\,;a,b,...=1,2,3$. The fixed coefficient $\nu$
is related to the constant Ricci
scalar curvature via $R=4n(n+2)\nu$ and therefore in four dimensions, $\nu=\Lambda/3$ with the standard
convention for the cosmological constant $\Lambda$ as in the previous section.

\subsection{Twistor space and complex contact structure}

A quaternion-K\"ahler manifold $(\cM,g,{\vec J})$ can be efficiently described by means of
its twistor space $\cZ$,  a $\CP$ bundle over $\cM$, whose connection is given
by the $SU(2)$ part $\vec p$ of the Levi-Civita connection on $\cM$.
We denote by $\pi$ the projection from $\cZ$ to $\cM$.
$\cZ$ admits a canonical integrable
complex structure and a (pseudo\footnote{'Pseudo' refers
to the fact that for negative scalar curvature $\Lambda < 0$, the metric \eqref{Z-metric}
is pseudo-Riemannian with signature $(2,4)$, whereas for $\Lambda>0$ it is positive definite.}-)
 K\"ahler-Einstein
metric \cite{MR664330}
\begin{equation}\label{Z-metric}
\de s^2_{\cZ}=
\frac{|Dt|^2}{(1+t{\bar t})^2}+\frac{\Lambda}{12}\,\de s_{\cal M}^2\, .
\end{equation}
Here $t$ is a complex stereographic coordinate on $\CP$,
and  $Dt$ is a one-form of Dolbeault type $(1,0)$ on $\cZ$,
\be
\label{defdz}
Dt \equiv  \de t + p^+ -\I p^3 \,t + p^-\, t^2\, ,
\ee
where we define $v^\pm\equiv -\half(v^1\mp \I v^2)$ for any real vector $\vec v$.
Under $SU(2)$ frame
rotations, $\vec p$ transforms as a $SU(2)$ connection while $t$ undergoes Moebius
transformations. More concretely, under the infinitesimal transformations
\begin{equation}
\delta t= \epsilon^+-\I \epsilon^3 t +\epsilon^- t^2\ ,\quad
\delta p^+=-\I\epsilon^3p^++\I \epsilon^+p^3-{\rm d}\epsilon^+\ ,\quad
\delta p^3 =2\I (\epsilon^-p^+-\epsilon^+p^-)-{\rm d}\epsilon^3\ ,
\end{equation}
$Dt$ transforms homogeneously and the first term
in \eqref{Z-metric} is gauge invariant.

While $Dt$ is only defined up to a multiplicative factor, its kernel is globally well-defined
and endows $\cZ$ with a canonical complex contact structure \cite{MR664330}.
As explained in \cite{Alexandrov:2008nk,Alexandrov:2008gh},  the contact structure $\CX$
descends naturally from the Liouville one-form on the Swann bundle (or in physics parlance,
hyperk\"ahler cone) $\cS$.
In practice,
it is convenient to use an open covering $\cU_i$ of $\cZ$, and represent the complex contact structure
by an holomorphic one-form $\cX\ui{i}$ defined on each patch $\cU_i$, such that $\cX\ui{i}$
and $\cX\ui{j}$ differ on the overlap $\cU_i \cap \cU_j$ by multiplication by a
non-vanishing holomorphic function. While the  $(1,0)$ form $Dt$ is not holomorphic
(i.e. $\bar\partial$-closed), it lies in the kernel of the complex contact distribution,
so on each patch (after suitable refinement) there exists a complex valued function $\Phi^{[i]}$
such that \cite{Alexandrov:2008nk,Alexandrov:2008gh}\footnote{The factor of $2/t$ in \eqref{c-pot}
is purely conventional.}
\begin{equation}
\label{c-pot}
\mathcal {X}^{[i]} = 2\, {\rm e}^{\Phi^{[i]}} \,\frac{Dt}{t}\, .
\end{equation}
The function $\Phi\ui{i}$, which we call the contact potential, is non-holomorphic on $\cZ$
but its restriction to each fiber is, so we denote it as $\Phi^{[i]}(x^\mu,t)$ where $x^\mu$
are coordinates on $\cM$. It is of course subject to the same ambiguity as the contact
one-form $\cX\ui{i}$, namely
\be
\label{ambi}
(\cX\ui{i},  \Phi\ui{i} ) \sim \left( f_i^2 \cX\ui{i}, \Phi\ui{i} + \log(  f_{i}^{2}) \right) \, ,
\ee
where  $f_i^2$  is a holomorphic function on $\cU_i$. Moreover, on the overlap of two
different patches, there must exist holomorphic gluing functions $f_{ij}^2$ such that
\be
\label{glue2}
\CX\ui{i} =   f_{ij}^{2} \, \CX\ui{j} ,
\qquad
\Phi^{[i]}-\Phi^{[j]}  =  \log f_{ij}^{2}\, .
\ee
Combining \eqref{dsMZ}, \eqref{c-pot} and
\eqref{Z-metric},  it is now obvious that the contact potential determines a
\kahler potential on $\cZ$ via
\be
K\ui{i}=\log \(2\,\frac{1+t\bar t}{|t|}\)+\Re\Phi\ui{i} (x^\mu,t)\, ,
\label{Kpotential}
\ee
and that the ambiguity \eqref{ambi} amounts to \kahler transformations.
Moreover, the Swann bundle $\cS$, with \hk
metric  \cite{deWit:2001dj,Neitzke:2007ke}
\be
\de s^2_\cS = {\rm e}^{\lambda^{[i]}+\bar\lambda^{[i]}+K\ui{i}}
\left(  | \de\lambda^{[i]} + \pa K\ui{i}|^2 + \de s^2_\cZ \right) =
|D\pi|^2 + |\pi|^2 \de s^2_\cM\ ,
\ee
where  $\pi^{A'}$ parametrizes the  $\IC^2/\IZ^2$ fiber of $\cS$,
$D\pi^{A'}=\de \pi^{A'} + {p^{A'}}_{\! B'} \pi^{B'}$ is the covariant
derivative of $\pi^{A'}$, and ${\rm e}^{\lambda\ui{i}}$ is a local coordinate on the
$\cO(-1)$ line bundle over $\cZ$, can be reconstructed from the contact potential
$\Phi\ui{i}$  via
\be
\label{pisec}
\pi^{A'} = {\rm e}^{\lambda^{[i]}-\frac12 \Phi\ui{i}}
\begin{pmatrix} t^{1/2} \\ t^{-1/2}
\end{pmatrix}\ .
\ee
This relation was first obtained in  the case of toric \qk manifolds in \cite{Neitzke:2007ke},
but holds in fact in general, as can be seen by  combining Eqs. 2.14,20,22,77,79 in \cite{Alexandrov:2008nk}.

\subsection{Darboux coordinates and transition functions}

By Darboux theorem, any contact structure is locally trivial. Thus, possibly after
refining the covering, it is possible to choose complex coordinates on $\cU_i$
such that the complex contact form becomes canonical, i.e.
\begin{equation}\label{c-form}
\mathcal {X}^{[i]} \equiv {\rm d}\alpha^{[i]} +\xi^{[i]} {\rm d}{\tilde \xi}^{[i]}\, .
\end{equation}
On the overlap $\cU_i\,\cap \,\cU_j$, the two coordinate systems must be related by
a complex contact transformation. In analogy with classical mechanics\footnote{The
transition function $\hHij{ij}$ is related to the standard Hamilton function $\hSij{ij}$ via
$\hSij{ij} = \ai{j} + \xii{i} \,  \txii{j} - \hHij{ij}( \xii{i},\txii{j},\ai{j})$. In particular, $\hHij{ij}=0$
for the identity map.},
such transformations may be parametrized  via a transition function
 $\hHij{ij}(\xii{i},\txii{j},\ai{j})$ of the ``initial position''
$\xii{i}$, ``final momentum" $\txii{j}$ and ``final action" $\ai{j}$
such that, on $\cU_i\cap \cU_j$,
\be
\begin{split}
\xii{j} = &  \xii{i} + T^{[ij]},
\qquad
\txii{j} =  \txii{i}
 + \tilde{T}^{[ij]} ,
\qquad
\ai{j} =  \ai{i}
 + \tilde{T}^{[ij]}_\alpha ,
\end{split}
\label{QKgluing}
\ee
where we defined
\be
\begin{split}
\label{Tfct}
& T^{[ij]} \equiv
 -\p_{\txii{j} }\Hij{ij}
+\xii{j} \, \p_{\ai{j} }\Hij{ij}  ,
\\
\tilde{T}^{[ij]} \equiv &
 \p_{\xii{i} } \Hij{ij} ,
\qquad
\tilde{T}^{[ij]}_\alpha\equiv
\Hij{ij}- \xii{i} \p_{\xii{i}}\Hij{ij}.
\end{split}
\ee
The transition functions are subject to consistency conditions on triple overlaps
$\cU_i\cap \cU_j\cap \cU_k$, gauge invariance on $\cU_i$ and reality conditions
as described in \cite{Alexandrov:2008nk,Alexandrov:2008gh}.
The original \qk metric on $\cM$ may be recovered from this holomorphic data
by constructing the ``real twistor lines", i.e. by solving the gluing conditions
\eqref{QKgluing} and expressing the complex coordinates $\xii{i},\txii{i},\ai{i}$
on $\cZ$ in terms of the fiber coordinate $t$ and the base coordinates $x^\mu$.
The $SU(2)$ connection $\vec p$ can then be read off by Taylor expanding
$\cX\ui{i}$ around any point $t=t_i$ in $\cU_i$ and using \eqref{c-pot}, \eqref{defdz}.

\subsection{Integral equations  for the twistor lines \label{sectzero}}

The Darboux coordinates $\xii{i},\txii{i},\ai{i}$, viewed as functions of the fiber coordinate $t$
for a fixed point $x^\mu\in \cM$, are usually required to be analytic inside the patch $\cU_i$.
Upon analytic continuation outside $\cU_i$, one generically encounters singularities.
As demonstrated in \cite{Alexandrov:2008nk,Alexandrov:2008gh}, expanding around
such a singular point can significantly simplify the retrieval of the metric on $\cM$. In particular, it is
useful to relax slightly the condition of regularity of the Darboux coordinates and  allow for
a simple pole in the Darboux coordinate $\xi$ at $t=0$ and at the antipodal point $t=\infty$,
while  $\txi$ and $\alpha$ can have a logarithmic singularity at these points.
This singularity structure emerges naturally from the superconformal quotient of toric \hk
manifolds, and continues to hold for a class of finite deformations, though we do not
expect it to be generally valid. As will become clear in Section 3, the advantage of such a choice is that
the section $t(x^\mu)=0$ is the vanishing locus of a holomorphic function, namely $1/\xi=0$,
and therefore leads to a (local) integrable complex structure on $\cM$.

To keep our notations consistent with \cite{Alexandrov:2008nk,Alexandrov:2008gh},
we assume that $t_0=0$ belongs to the ``north pole" patch $\cU_+$.
According to our assumption, the Darboux coordinates must therefore behave
at $t=0$ as \cite{Alexandrov:2008nk}
\be\label{behap}
\begin{split}
\xii{+} &= \xii{+}_{-1}\, t^{-1} + \xii{+}_{0}
+ \xii{+}_{1}\,  t +\cO( t^2)\, ,  \qquad
\\
\txii{+} &=  \tc \log t + \txii{+}_{0}
+ \txii{+}_{1}\, t+\cO( t^2)\, ,
\\
\ai{+} &= \ac \log t + \tc \, \xii{+}_{-1} \,t^{-1}
+ \ai{+}_0+ \ai{+}_1 \,t+ \cO( t^2)\, ,
\\
\Phi\ui{+}&= \phi\ui{+}_0 +  \phi\ui{+}_1 \, t +\cO( t^2) \, .
\end{split}
\ee
The coefficients $\ac$ and $\tc$ of the logarithmic singularity
are real numbers sometimes called  ``anomalous dimensions''.
Together with the transition functions, they completely specify
the twistor space and the original \qk manifold.

Assuming the behavior \eqref{behap} at $t=0$ and enforcing the reality conditions
described in \cite{Alexandrov:2008nk,Alexandrov:2008gh},
the gluing conditions \eqref{QKgluing}
can be rewritten as the following {\it exact}
integral equations \cite{Alexandrov:2009zh}
\bea\label{twist-lines}
\xii{i}
(\varpi,x^\mu)& =& A +
 \cR \(\varpi^{-1}- \varpi \)
+\frac12 \sum_j \oint_{C_j}\frac{\de\varpi'}{2\pi\I \varpi'}\,
\frac{\varpi'+\varpi}{\varpi'-\varpi}\, T^{[+j]}(\varpi') ,
\nonumber \\
\txi^{[i]}(\varpi,x^\mu)& = &\frac{\I}{2}\, B +
\half  \sum_j \oint_{C_j} \frac{\de \varpi'}{2 \pi \I \varpi'} \,
\frac{\varpi' + \varpi}{\varpi' - \varpi}
\, \tilde{T}^{[+j]}(\varpi')+  \tc \log \varpi ,
\label{txiqline}
\\
\ai{i}(\varpi,x^\mu)& = &\frac{\I}{2}\, B_\alpha +
\half  \sum_j \oint_{C_j} \frac{\de \varpi'}{2 \pi \I \varpi'} \,
\frac{\varpi' + \varpi}{\varpi' - \varpi}
\, \tilde{T}^{[+j]}_\alpha(\varpi')+  \ac \log \varpi
+\tc\cR\( \varpi^{-1} +\varpi\),
\nonumber
\eea
where $C_j$ is a contour surrounding the patch $\cU_j$,  and
$x^\mu=\{\cR, A, B, B_\alpha\}$ are real parameters which can be used
as coordinates on $\cM$. These equations
generalize the  Riemann-Hilbert problem discussed in \cite{Gaiotto:2008cd}
in the context of \hk geometry, and determine the
twistor lines as a formal power series in $H$, which we assume
to be convergent.

Having determined the twistor lines in this fashion,
the contact potential can obtained as
\be
\Phi^{[i]}(t,x^\mu)=\phi(x^\mu)-\frac12 \sum_j \oint_{C_j} \frac{\de \varpi'}{2 \pi \I \varpi'}
\,\frac{\varpi' + \varpi}{\varpi' - \varpi}\, \log\(1- \p_{\ai{j} }\Hij{+j}(\varpi')\)\ ,
\label{solcontpot}
\ee
where the real function $\phi(x^\mu)$ (the real part of $\Phi^{[+]}$ evaluated at $t=0$)
is determined in terms of the base coordinates via
\be
{\rm e}^{\phi(x^\mu)}=\frac{\frac{\cR}{2} \sum_j\oint_{C_j}\frac{\de\varpi}{2\pi\I \varpi}
\(\varpi^{-1} -\varpi  \)\p_{\xii{+}}\hHij{+j}+ \tc A + \ac}
{2\cos\[\frac{1}{4\pi} \sum_j\oint_{C_j}\frac{\de\varpi}{\varpi}\, \log\(1-\p_{\ai{j}}\Hij{+j} \)\]}.
\label{contpotconst}
\ee
Note that \eqref{solcontpot} reduces to $\Phi\ui{i}(t,x^\mu)=\phi(x^\mu)$ in the special
case where the transition functions $H^{[ij]}$ are independent of the action variable $\ai{j}$.
This is the relevant case when $\cM$ admits one isometry, as discussed later in Section 4.

Following the method outlined in the previous subsection,
one may extract the $SU(2)$ connection by expanding around $t=0$,
\be
\begin{split}\label{connection}
p_+ &=\frac12\, {\rm e}^{-\phi\ui{+}_0}
\left( \xii{+}_{-1}  \de\txii{+}_{0}
+ \tc \de\xii{+}_{-1}
\right)\, ,
\\
p_3 &= \frac{\I}{2}\, {\rm e}^{-\phi\ui{+}_0} \left( \de\ai{+}_0 +
\xii{+}_{0}  \de\txii{+}_{0} +
\xii{+}_{-1}  \de\txii{+}_{1}  \right)
- \I \phi\di{+}^1 p_+\, ,
\end{split}
\ee
where  $\phi\di{+}^0, \phi\di{+}^1$ can be expressed in terms of the
Laurent coefficients of the contact twistor lines,
\be
\label{contpot}
\begin{split}
{\rm e}^{\phi\ui{+}_0} =&  \frac12 \left( \xii{+}_{-1} \txii{+}_{1}
+ \ci{+}_\txi \xii{+}_{0} +   \ac \right)\, ,
\\
\phi\ui{+}_1 = & \frac12 \,{\rm e}^{-\phi\ui{+}_0}
\left( \ai{+}_1 + 2 \xii{+}_{-1}  \txii{+}_{2}
+  \xii{+}_{0}  \txii{+}_{1} + \tc \xii{+}_{1}  \right)\, .
\end{split}
\ee
By expanding the holomorphic one-forms
$\de\xii{+},\ \de\txii{+}$ and $\de\alpha$
around $\varpi=0$ and projecting along the base $\cM$, one may also
obtain a basis of (1,0)-forms on  $\cM$ for the
quaternionic structure $J_3$. Finally,
by using the second equation in \eqref{ompp}, one may further compute the
quaternionic 2-forms, and finally obtain the \qk metric on $\cM$.

\subsection{Constraints on the \kahler potential \label{seccons}}

As explained in \cite{deWit:2001dj}, the existence of a complex contact structure
implies certain constraints on the \kahler potential $K(u^i,\bar u^{\bar \imath})$.\footnote{In this subsection
we drop the patch index $[i]$, but introduce complex coordinates $u^i, i=1,2,3$ on $\cZ$, hoping
that the use of the same letter will not confuse the reader.} Let us define the
holomorphic two-form $\omega_{ij}$, Reeb vector $\cY^i$ and (antisymmetric)
bi-vector $\hat\omega^{ij}$ via
\be
\omega_{ij}= -\p_{[i}\cX_{j]}\, ,
\qquad
\cX_i\cY^i =  1\, ,
\qquad
\omega_{ij}\cY^j =  0\, ,
\qquad
\hat\omega^{ik}\omega_{kj} =  -\delta^{i}_j+\cY^i\cX_j\, .
\label{constrT}
\ee
Then the constraints that $\cZ$ is the twistor space of a \qk manifold can
be summarized by the three equations
\be
\label{defomm}
\cY^i=\cX^i {\rm e}^{-2K}-\hat \omega^{ij}K_j \, ,
\qquad
\hat\omega^{ij}=\(\omega^{ij}+K^{i} \cX^{j} - K^{j} \cX^{i} \){\rm e}^{-2K}\ ,
\ee
\be
\det K_{i\bar \jmath} = \frac{1}{4}\, {\rm e}^{-4 K}\, ,
\label{MAeq}
\ee
where the indices are raised and lowered with the metric $K_{i\bar \jmath}$ and its inverse $K^{i\bar \jmath}$.
In particular, the requirement that the metric on $\cZ$ should be K\"ahler-Einstein is the
the Monge-Amp\`ere-Liouville equation \eqref{MAeq}. Moreover, it follows from \eqref{constrT}
that $\hat\omega^{ij} \cX_j =  0$, and from \eqref{defomm} that $\cX_i \cX^i ={\rm e}^{-2K}$: the
K\"ahler potential on $\cZ$ is thus obtained as the logarithm of the norm of the contact structure.

In Darboux coordinates, the above
constraints can be made more explicit. Indeed,
the conditions \eqref{constrT} uniquely specify
\be
\omega = -\de\xi \wedge \de\txi\, ,
\qquad
\cY = \pa_\alpha\, ,
\qquad
\hat\omega =
2\xi \pa_\xi \wedge \pa_\alpha - 2 \pa_\xi\wedge \pa_{\txi}\, ,
\ee
while the constraints \eqref{defomm}  reduce to the following three equations,
\be
\begin{split}
K_\xi= & \,\frac12 \(K^{\txi\balp}+\bxi K^{\txi\btxi}\){\rm e}^{-2K},
\\
K_{\txi}-\xi K_\alpha= &\, -\frac12 \(K^{\xi\balp}+\bxi K^{\xi\btxi}\){\rm e}^{-2K},
\\
{\rm e}^{2K}= &\, K^{\alpha\balp}+\xi K^{\txi\balp}+\bxi K^{\alpha\btxi}+|\xi|^2 K^{\txi\btxi}\, ,
\end{split}
\label{constrK}
\ee
to be supplemented by the Monge-Amp\`ere Equation \eqref{MAeq}. These constraints
will be useful in the next section when deriving the master equation \eqref{master} from
the twistor space.

\section{Przanowski metrics from twistor space}
\label{sec_PrfromT}

Having recalled generalities about \qk spaces in the previous section, we now come
to the main goal of this note, which is to clarify the geometric nature of the Przanowski
metric \eqref{metPrz} and to understand when different solutions of the master equation
\eqref{master} lead to \qk metrics that are related by diffeomorphisms. All of this section
holds independent of the assumptions made in section \ref{sectzero}, except when
we explicitly use some properties of the twistor lines for the $t=0$ slice leading to \eqref{relhPhi}.
At the end of this section, we also show how the full twistor space can be recovered
from a single ``tomogram", or a Przanowski solution, by solving
a recursive system of differential equations. This system  simplifies considerably
in the presence of isometries. In particular, the contact potential can be found
exactly, see \eqref{rescph}, though the twistor lines must still be obtained recursively.

\subsection{From self-dual Einstein spaces to Przanowski metrics}
\label{subsec_fromTtoQK}

As already mentioned in the introduction, the metric \eqref{metPrz}
is manifestly hermitian with respect to a local integrable complex structure $J$
with complex coordinates $x^\mu=(z^1,z^2,\bz^1,\bz^2)$.
To show that the metric  \eqref{metPrz} is
\qk,  it suffices to check that
the conditions \eqref{ompp} are satisfied  for
\begin{equation}
p^+={\rm e}^{h/2}{\rm d}z^2\, ,
\qquad
p^3=\frac{i}{2}(\partial -{\bar \partial})h\, ,
\qquad
p^-=(p^+)^*\, ,
\label{su2-conn}
\end{equation}
\begin{equation}
\label{2-forms}
\omega^+=\frac{6}{\Lambda}\,{\rm e}^{h/2}h_1\,{\rm d}z^1\wedge {\rm d}z^2\, ,
\qquad
\omega^3 =2\I \,g_{\alpha\bar\beta}\,{\rm d}z^\alpha \wedge {\rm d} z^{\bar \beta}\, ,
\qquad
\omega^-= (\omega^+)^*\, ,
\end{equation}
where
\be
g_{\alpha\bar\beta}=-\frac{3}{\Lambda}\(\p_{\alpha}\p_{\bar\beta}h+
2\delta_{\alpha}^2\delta_{\bar \beta}^{\bar 2}\, {\rm e}^h \) ,
\label{gPrz}
\ee
and $\pa=\de z^1 \pa_1 + \de z^2 \pa_2$ is the Dolbeault operator associated to $J$.
A straightforward computation shows that  \eqref{ompp} indeed holds
provided $h$ satisfies the master equation \eqref{master}.  Eqs. \eqref{su2-conn} and
\eqref{2-forms}  were obtained in a different local $SU(2)$ frame in \cite{Looyestijn:2008pg}.
One virtue of the above choice is that
$\omega_+$ and $p_+$ are of Dolbeault type (2,0) and (1,0)
with respect to $J$.  This would
of course also continue to hold after $U(1)$ frame rotations
\begin{equation}
\label{u1}
\omega^+ \rightarrow {\rm e}^{\I \theta}\,\omega^+\, ,
\qquad
\omega^3 \rightarrow \omega^3\, ,
\qquad
p^+\rightarrow {\rm e}^{\I \theta}\,p^+\, ,
\qquad
p^3\rightarrow p^3 +{\rm d}\theta\, ,
\end{equation}
for any real function $\theta(x^\mu)$.

To show that any self-dual Einstein metric can be locally cast into the Przanowski Ansatz,
we start by choosing a linear combination $J=n_a(x^\mu) J^a\,; a=1,2,3$ of the quaternionic structures, where
$\vec n$ is a unit three-vector, such that $J$ is (locally) an integrable complex structure.
As discussed in the next subsection, there are infinitely many choices of $n_a(x^\mu)$
corresponding to choices of complex submanifolds in the twistor space $\cZ$. Upon
performing a suitable $SU(2)$ frame rotation, there is no loss of generality in assuming
that $J=J_3$. The corresponding quaternionic form $\omega^+$ and connection
$p^+$ are then of Dolbeault type (2,0) and (1,0) with respect to $J$. One may now perform
a suitable holomorphic change of variable such that $p^+$ is proportional to the
holomorphic one-form $\de z^2$, i.e.
\begin{equation}\label{def-z2}
p^+= {\rm e}^{\cA}\, {\rm d}z^2\, ,
\end{equation}
for some complex-valued function $\cA(x^\mu)$. Clearly, $z^2$ is ambiguous up
to holomorphic changes of variables of the form \eqref{diff-equiv}, whereby $\cA$
changes as $\cA\rightarrow \cA-\log g_2(z^2)$. The conditions \eqref{ompp},
written in component as
\begin{equation}\label{d-omega}
{\rm d}\omega^+ +\I p^+\wedge \omega^3 - \I  p^3\wedge \omega^+=0\, ,
\qquad
{\rm d}\omega^3 -2\I p^+\wedge \omega^- +2 \I p^-\wedge \omega^+=0
\end{equation}
for the curvature two-forms, and
\begin{equation}\label{su2-curv}
{\rm d}p^++\I p^+\wedge p^3=\frac{\Lambda}{6}\,\omega^+\, ,
\qquad
{\rm d}p^3-2\I p^+\wedge p^-=\frac{\Lambda}{6}\,\omega^3\, ,
\end{equation}
then uniquely specify
\begin{equation}
p^3=\I \(\partial_\alpha {\bar \cA}\, {\rm d}z^\alpha - \partial_{\bar \alpha}
\cA\,{\rm d}{\bar z}^\alpha\)\ ,\qquad
\omega^+=\frac{12}{\Lambda}\,{\rm e}^\cA\partial_{z^1}({\rm Re}\,\cA)\,{\rm d}z^1\wedge {\rm d}z^2\, .
\label{connect_A}
\end{equation}
Computing $\omega^3$  from the second equation in \eqref{d-omega} and multiplying
with the complex structure $J^3$ then leads to the metric \eqref{metPrz}, where the
Przanowski function is identified with the real part of $\cA$,
\begin{equation}
h = \cA + \bar \cA \, .
\end{equation}
In particular, under holomorphic changes of variables of the form \eqref{diff-equiv},
$h$ changes as prescribed in  \eqref{equiv-h}.
On the other hand, the imaginary part of $A$ can be shifted away after a $U(1)$ rotation,
since $\cA\to \cA+\I \theta$ induces the rotation \eqref{u1} on the connection and quaternionic
forms. The choice made in Appendix C.2 of \cite{Looyestijn:2008pg} corresponds to
$\cA-{\bar \cA}=\log (h_{\bar 1}/h_1)$, while the one in \eqref{su2-conn} corresponds to
$\cA-{\bar \cA}=0$.

\subsection{From twistor spaces to Przanowski metrics \label{sechat}}

We now discuss how to obtain the  Przanowski metric \eqref{metPrz} in a top-down approach,
starting from the twistor space $\cZ$. Recall that for any \kahler-Einstein space $\cZ$ equipped with a
complex contact structure $\cX$, the quadratic differential
\be
\label{dszx}
{\rm d}s^2_{\cM}\equiv \frac{12}{\Lambda} \left(
{\rm d}s^2_{\cZ}-{\rm e}^{-2K}|\cX|^2\right)\ ,
\ee
has a two-dimensional kernel and signature (4,0). Its restriction to any real-codimension 2
(local) submanifold $\cC$ transverse to the contact distribution, i.e. such that
$\cX\vert_\cC\neq 0$, produces local
self-dual Einstein metrics on $\cM$ that are diffeomorphic to each other. Suppose
now that $\cC$ is a complex submanifold of $\cZ$, given by the vanishing of
some holomorphic function $\cC(u^i)=0$ on $\cZ$. The first term in \eqref{dszx}
restricts to a K\"ahler metric, locally defined on $\cM$, whose K\"ahler potential $K_{\cC}$ is the
restriction of the K\"ahler potential $K$ on $\cZ$ to the submanifold $\cC$.
As for the second term, two possibilities arise:
\begin{enumerate}
\item[i)] either the two-form $\de\cX\vert_{\cC}=0$ vanishes on $\cC$; in this case there
exists a holomorphic function $z^2$ such that $ \cX\vert_{\cC}=\de z^2$; since $z^2$
is non-constant, it can be taken as a complex coordinate on an open subset in $\cC$, and
supplemented by a second coordinate $z^1$ on $\cC$ such  that $\de z^1 \wedge \de z^2\neq 0$;
\item[ii)] or the two-form $\de\cX\vert_{\cC}$ is non-degenerate on $\cC$; by Darboux's theorem,
there exists complex coordinates $z^1, z^2$ on an open subset in $\cC$ such that
$ \cX\vert_{\cC}=z^1 \de z^2$;
\end{enumerate}
Both cases can be treated simultaneously by assuming  that
\be
\cX\vert_{\cC}={\rm e}^{F_{\cC}(z^1,z^2)}\, \de z^2
\label{redX}
\ee
on an open subset of $\cC$, coordinatized by complex coordinates $z^1, z^2$.
This covers cases i) and ii) above, with ${\rm e}^{F_{\cC}}=1$ and ${\rm e}^{F_{\cC}}=z^1$
respectively, but allows for more general coordinate choices.
Identifying
\be
h=-2K_{\cC}+F_{\cC}+\bar F_{\cC}\ ,
\label{identh}
\ee
we see that the metric \eqref{dszx} reduces to the Przanowski form; in
particular, $h$ defined in \eqref{identh} must be a solution of the master
equation \eqref{master}. In Appendix A, we show how this
follows from the constrains of twistor geometry discussed in
Section \ref{seccons} for the special choice $\cC(u^i)=\xi$, but the derivation can be easily
generalized to any slice.
Moreover, the quaternionic structures \eqref{2-forms} can
also be obtained from the twistor space using \cite{deWit:2001dj}
\be
 \omega^3_{\alpha\bar\beta} =
 -\I\left( K_{\alpha\bar\beta} - {\rm e}^{-2K} \cX_\alpha \cX_{\bar\beta}\right)\, ,
\qquad
 \omega^+_{\alpha\beta}={\rm e}^{-K}\, (\omega_{\alpha\beta} + K_\alpha
 \cX_\beta- K_\beta \cX_\alpha)\, .
\ee

As in \eqref{def-z2}, Eq. \eqref{redX} determines $z^2$ up
to holomorphic changes of variables of the form \eqref{diff-equiv}, whereby $F_{\cC}$
changes by $F_{\cC}\rightarrow F_{\cC}-\log g_2(z^2)$, reproducing the
holomorphic ambiguity of $h$ in \eqref{equiv-h}. More significantly
however, changing  the complex submanifold $\cC$ produces
new solutions of the master equation \eqref{master} which describe
the same metric $\cM$ but are related to each other by non-holomorphic
changes of variables, as we shall discuss in Section \ref{secnonh}.

In practice, it is useful to choose Darboux
coordinates $u^i=(\alpha,\xi,\txi)$ on $\cZ$, such that the contact structure
$\cX$ takes the form \eqref{c-form}, and parametrize $\cM$ by solving
the condition $\cC(u)=0$ for one of the coordinates $u^i$ in terms of the other.
In the remainder of this subsection, we illustrate this process for several choices of
parametrizations.

\subsubsection{$\hat\xi$-parametrization}
\label{subsubsec_xi}

Suppose the submanifold $\cC$ can be described locally as $\xi=\hat\xi(\alpha,\txi)$,
where $\hat\xi$ is any holomorphic function of $\alpha,\txi$. The condition \eqref{redX}
then reduces to the three conditions
\be
\p_{\txi} z^2={\rm e}^{-F_{\hat\xi}}\, \hat\xi\ ,\qquad \p_\alpha z^2={\rm e}^{-F_{\hat\xi}}\ ,\qquad
 \(\p_{\txi}-\hat\xi\p_\alpha\) F_{\hat\xi} +\p_\alpha\hat\xi=0\ .
\label{Fcoorxi}
\ee
for the functions $F_{\hat\xi}(\alpha,\txi)$ and $z^2(\alpha,\txi)$.
The last condition is in fact the integrability condition for the first and second.
$z^1(\alpha,\txi)$ can be chosen arbitrarily in such a way that  the Jacobian
${\p (z^1,z^2)}/{\p (\alpha,\txi)}$ is non-zero.
Different solutions  will be  related by holomorphic changes of variables
of the form  \eqref{diff-equiv}.

A particular class of submanifolds $\xi=\hat\xi(\alpha,\txi)$
where \eqref{Fcoorxi} can be solved explicitly are those of the form
$\xi=\pa_{\txi} W(\txi)$, where $W$ is an arbitrary holomorphic function
of $\txi$: in this case, one may choose
\be
\label{Leg-slice}
z^1 = \txi\, ,
\qquad
z^2 = \alpha + W(\txi)\, ,
\qquad
F_{\hat\xi}= 0\, ,
\ee
leading to the following infinite class of solutions of Przanowski equation
\be
\label{hW}
h_W(z^1,z^2) = - 2 K\( W'(z^1), z^1 , z^2 - W(z^1)\)\, ,
\ee
where $W'$ denotes the derivative of $W$.
The case $W=0$ allows for a particularly simple derivation of the
master equation, as outlined in Appendix A.

\subsubsection{$\hat\alpha$-parametrization}

Suppose now that the submanifold $\cC$ is described locally as $\alpha=\hat\alpha(\xi,\txi)$.
The condition \eqref{redX} now reduces to the three conditions
\be
\pa_\xi z^2={\rm e}^{-F_{\hat\alpha}}\p_\xi\hat\alpha,
\qquad
\pa_{\txi} z^2={\rm e}^{-F_{\hat\alpha}}\(\xi+\pa_{\txi}\hat\alpha\).
\label{defeqz2}
\ee
\be
\pa_\xi \hat\alpha\, \pa_\txi F_{\hat\alpha}- \pa_\txi \hat\alpha\, \pa_\xi F_{\hat\alpha}
=\xi\pa_\xi F_{\hat\alpha} -1.
\label{eqdeff}
\ee
Again, the third equation is the integrability condition for the first two.

\subsubsection{$t=0$ slice \label{sect0}}

A particular convenient choice of submanifold $\cC$ is the special locus $t(x^\mu)=0$,
introduced in Section \ref{sectzero}, where one of the Darboux coordinates $\xi$
diverges linearly. This may be viewed as a special case of the $\hat\xi$-parametrization,
given formally by  $\hat\xi=\infty$. This complex submanifold  was used in the
superconformal quotient constructions
of \cite{deWit:2001dj,Rocek:2005ij,Alexandrov:2007ec}, and is  particularly
convenient for reconstructing the twistor space from the
Przanowski function $h$ in the presence of one isometry, as will become apparent
in Section \ref{subsec_fromPtoT}.

Since the contact potential $\Phi$
is assumed to be regular on the section $t=0$, the \kahler potential \eqref{Kpotential} diverges
at that point.
This is repaired by a \kahler transformation
\be
K\mapsto K'=K-\log|\xi|=\Re\Phi-\log \frac{|t\xi|}{2(1+t\bt)}\, ,
\ee
combined with a rescaling of the contact form \eqref{c-form},
\be
\cX\mapsto \cX'=\xi^{-1}\cX=\de\txi+\xi^{-1}\de \alpha\, .
\ee
Now the limit $t=0$ is regular and yields
\be
K'_{t=0}=\phi(x^\mu)-\log\frac{ |\xii{+}_{-1}|}{2}\, ,
\qquad
\cX'\vert_{t=0}=\de (\txii{+}_0+\tc \log\xii{+}_{-1} )\, ,
\label{reductX}
\ee
where $\xii{+}_{-1}, \txii{+}_0$ are the Laurent coefficients appearing in \eqref{behap}.
Defining the complex coordinates $z^1, z^2$ as
\be
z^1=\ai{+}_0+\ac\log \xii{+}_{-1}-\tc A\, ,
\qquad
z^2=\txii{+}_0+\tc\log \xii{+}_{-1}\, ,
\label{defcoor}
\ee
such that $F_{t=0}=0$, we can finally cast the metric into the
Przanowski Ansatz with
\be
h=-2K'_{t=0}=-2\phi(x^\mu)+2\log\frac{\cR}{2}.
\label{relhPhi}
\ee

\subsubsection{Relating different parametrizations}
\label{subsec_reldifslice}

While each of the parametrizations above have their own virtues, it turns out that they
can all be mapped to each other (locally) by a suitable contact transformation. In particular,
the slice $\xi=\hat\xi(\txi,\alpha)$ can be mapped to $\xi'=\infty$ via
the contact transformation
\bea
\xi'&=&  -{\rm e}^{F_{\hat\xi}}/(\xi-\hat\xi),
\nn\\
\txi'+\tc\log\xi' & = &z^2(\txi,\alpha)+\ac (\xi-\hat\xi) {\rm e}^{-F_{\hat\xi}},
\label{coortr_xi}\\
\alpha'-\tc\xi'+\ac\log\xi' & =& -\txi,
\nn
\eea
where $F_{\hat\xi}$ satisfies the differential equation \eqref{Fcoorxi}.
Indeed, it is readily checked that $\cX' /\xi' = {\rm e}^{-F_{\hat\xi}} \cX$.

Similarly, the slice $\alpha=\hat\alpha(\xi,\txi)$ can be mapped to the slice $\xi'=\infty$
by the following complex contact transformation,
\bea
\xi' &=&1/ \(\alpha-\hat\alpha\)\, ,
\nn
\\
\label{coortr_al}
\txi'+\tc\log\xi' &=&z^2(\xi,\txi)+\({\rm e}^{-F_{\hat\alpha}}-\ac\)\(\alpha-\hat\alpha\),
\\
\alpha'-\tc\xi'+\ac\log\xi'& =& -{\rm e}^{-F_{\hat\alpha}}\, ,
\nn
\eea
such that $\cX' /\xi'= {\rm e}^{-F_{\hat\alpha}} \cX$. Here
$F_{\hat\alpha}$ is the function defined by the differential equation \eqref{eqdeff}.
By combining these two transformations, one can dispose of  the singular
slice $\xi'=\infty$ and relate the $\hat\xi$
and $\hat\alpha$ parametrizations directly,
\bea
\alpha-\hat\alpha &=&  -{\rm e}^{-F_{\hat\xi'}} (\xi'-\hat\xi'),
\nn\\
z^2(\xi,\txi) + {\rm e}^{-F_{\hat\alpha}}\(\alpha-\hat\alpha\)
& = &z'^2(\txi',\alpha'),
\label{coortr_xial}\\
{\rm e}^{-F_{\hat\alpha}} & =& \txi'\ ,
\nn
\eea
such that ${\rm e}^{-F_{\hat\alpha}} \cX = {\rm e}^{-F_{\hat\xi'}} \cX'$.

\subsection{Isometric solutions and non-holomorphic coordinate changes\label{secnonh}}

As indicated previously, different choices of submanifold $\cC$ lead to different Przanowski
representations of the same metric, where different (local) integrable complex structures
are manifest. There is however a clear way to relate different representations, depicted
in Figure \ref{fig1} on page \pageref{fig1}: let
\be
\begin{split}
i&: (z^1,z^2)\in V\to \cC \subset \cZ
\\
i' &: (z'^1,z'^2)\in V\to \cC' \subset \cZ
\end{split}
\ee
be the two submersions from open subsets $V,V'$ in $\IC^2$ into the submanifolds $\cC,\cC'$,
corresponding to two different Przanowski representations. Consider now
\be
\label{zzp}
z\in V \mapsto (i')^{-1} \left[ \pi^{-1}[ \pi (i (z) ] \cup \cC'\right] \ \in V' \, ,
\ee
where $\pi$ is the projection from $\cZ$ to $\cM$. In writing \eqref{zzp}, we have restricted
the open subsets $V,V'$ such that the twistor line above the point $\pi (i (z))$ intersects
$\cC'$ in one and exactly one point. Eq. \eqref{zzp} provides the map between the two
coordinate systems $(z^1,z^2)$ and $(z'^1,z'^2)$. Because the projection $\pi$ is non-holomorphic,
this map is also non-holomorphic. Of course, if $i(z)$ belongs to intersection $\cC \cap \cC'$, the
map reduces to $(i')^{-1}\circ i$, so becomes holomorphic, but this generically happens only
on a real-codimension 2 subspace of $V,V'$.

To illustrate this prescription, we now determine the change of variables between the
``special slice" $t=0$, corresponding to $\xi=\infty$, and a nearby slice parameterized by
\be
\hat\xi^{-1}=\p_\alpha W(\txi,\alpha)\, ,
\label{hatxiinv}
\ee
where $W$ is considered to be infinitesimal.  For simplicity, we
assume that the anomalous dimensions vanish. In this case
it is possible to solve equations \eqref{Fcoorxi} to first order in $W$ and obtain
\be
{\rm e}^{F_{\hat\xi}}\approx\frac{1-\p_{\txi}W   
}{\p_\alpha W},
\qquad
z^1\approx\alpha  
,
\qquad
z^2\approx \txi+W   
.
\ee
For $W=0$, the coordinates $z^\alpha$ and Przanowski function are given by
\be
z^1_{(0)}=\ai{+}_0,
\qquad
z^2_{(0)}=\txii{+}_0,
\qquad
h_{(0)}=-2\phi+2\log\frac{\cR}{2}.
\ee
For $W$ non-zero but vanishingly small,  perturbing around this solution leads to
\be
\begin{split}
z^1  &\approx z^1_{(0)}+\cR\, \ai{+}_1 \p_\alpha W,
\qquad\\
z^2 &\approx z^2_{(0)}+2 {\rm e}^{\phi^{[+]}_0}\p_\alpha W + W,\\
h& \approx h_{(0)}
-2\Re\[\p_{\txi}W+\cR \, \phi^{[+]}_1\, \p_\alpha W \]
\end{split}
\ee
which is manifestly non-holomorphic. In particular, the variation
\be
\label{delhnonhol}
\delta h = h-h_{(0)}- 2\Re\left( h_1 (z^1  -z^1_{(0)}) + h_2 (z^2 -  z^2_{(0)}) \right)
\ee
provides an
eigenmode of the linearized master equation \eqref{eqdPrz} outside the
class of trivial eigenmodes \eqref{pgt}. Unfortunately, it does not seem possible
to express $\delta h$ in terms of $h, W$ and their derivatives only.

\subsection{Lifting back to the twistor space \label{sec1iso}}
\label{subsec_fromPtoT}

In this section we address the inverse problem: given a solution $h$ of the Przanowski equation,
construct the corresponding twistor space, in particular provide the contact potential and
local Darboux coordinates. These are of course defined up to local
complex contact transformations (see Section 2.2). Since all local complex
submanifolds are equivalent up to a contact transformation, as explained
in Section \ref{subsec_reldifslice}, we can assume that $h$ is associated
to the  complex submanifold $t(x^\mu)=0$, along the lines of section \ref{sect0}.

Under this assumption, we can address this problem as follows: Eq. \eqref{c-pot} relates
the Darboux coordinates to the $SU(2)$ connection, which was
evaluated  in terms of the function $h$ in  \eqref{def-z2}, \eqref{connect_A}.
The Przanowski coordinates $(z^1,z^2)$ can in turn be related to
the Darboux coordinates using \eqref{defcoor}, where the coefficient $\xii{+}_{-1}$ is in general complex.
Moreover, one can use \eqref{relhPhi} in the form $\phi=-h/2+\log\frac{|\xii{+}_{-1}|}{2}$.
Comparing the resulting SU(2) connection with \eqref{connection} and using \eqref{contpot},
the first coefficients in the Laurent expansion of  $\xi$ at $t=0$ are found to be
\be
\xii{+}_{-1}=\frac{{\rm e}^{h/2+\I\theta}}{h_1}\, ,
\qquad
\xii{+}_0=\xii{+}_{-1}\phi^{[+]}_1+\frac{h_2}{h_1}\, ,
\label{resolgen}
\ee
while the constant part of the contact potential is
\be
\phi^{[+]}_0=-\log(2h_1).
\label{rescphgen}
\ee
Recall that $\theta=\Im \cA$ can be shifted arbitrarily by rotating the complex structures.
By choosing $\theta=-\frac{\I}{2}\,\log\frac{h_1}{h_{\bar 1}}$, we can ensure that
the coefficient $\xii{+}_{-1}=\cR$ is real,  as is necessary for  \eqref{txiqline} to hold.

To determine the higher order Laurent coefficients, one should solve a recursive
system of differential equations following from \eqref{c-pot}.
We demonstrate explicitly this in appendix \ref{ap_sol}
in the case when $\cM$ admits two commuting Killing vectors.
Such geometries can be described by Przanowski
solutions $h$ that depend only on $\Re z^\alpha$.
In this case, the Killing vector $\kappa=\I(\pa_1 - \pa_{\bar 1})$ lifts to a holomorphic
Killing vector $\kappa_\cZ=\I\pa_{\alpha}$, and therefore the transition functions $\hHij{ij}$
must not depend on $\alpha$. As explained in \cite{Alexandrov:2008nk},
this implies that the contact potential $\Phi\ui{i}$ is
the same in all patches and equal to the real function $\phi(x^\mu)$ so that from \eqref{rescphgen}
\be
\Phi\ui{i}=\phi=-\log(2h_1) .
\label{rescph}
\ee
On the other hand, the $\cO(2)$-valued moment map for the vector field
$\kappa'=\I(\pa_2 - \pa_{\bar 2})$ can be chosen to be one of the complex Darboux coordinates $\xi$,
given globally by
\be
\label{xiAR}
\xi=A+\cR\(t^{-1}-t\) ,
\ee
where $A=\xii{+}_0$ and $\cR$ are real and follow from \eqref{resolgen} with vanishing $\theta$
\be
\cR=\frac{{\rm e}^{h/2}}{h_1},
\qquad
\xii{+}_0=\frac{h_2}{h_1}.
\label{ressol}
\ee
Note that these identifications are valid even in the presence of only one isometry.
The remaining complex Darboux coordinates can be found order by order as expansions in $t$
by solving the differential equations \eqref{eqtxin}.

\section{Self-dual Einstein spaces with isometries}
\label{sec_isometr}

We now turn to self-dual Einstein spaces with isometries.
In Section \ref{secTod}, we review the relation between
self-dual Einstein spaces with one isometry and solutions of the Toda equation,
and map them to a particular
class of solutions of the Przanowski equation, following  \cite{Tod:2006wj} with a twist.
In Section \ref{sec_2iso}, we consider the case of two commuting isometries, and we show
how the Calderbank-Pedersen metrics can be brought into the Toda-form.

\subsection{Canonical complex structure and Toda equation \label{secTod}}

As shown by Tod \cite{MR1423177}, self-dual Einstein metrics
with one Killing vector field can be written in local coordinates $(\rho,z,\bar z, \psi)$
in the form
\be
\label{dstoda}
\de s_\cM^2 = -\frac{3}{\Lambda}\[
\frac{P}{\rho^2} \left( \de \rho^2 + 4 {\rm e}^ \todaQ \de z \de\bar z \right)
+ \frac{1}{P\rho^2}\,(\de \psi + \Theta )^2\] ,
\ee
where the isometry acts as a shift in the coordinate $\psi$.
Here, $ \todaQ$ is a function of $( \rho,z,\bar z)$,  $P\equiv 1- \haf\,\rho \pa_ \rho  \todaQ $,
and $\Theta$ is a one-form such that
\be
\label{dth}
\de \Theta = \I (\pa_z P \de z - \pa_{\bar z} P \de \bar z)\wedge \de  \rho
- 2\I\, \partial_\rho(P {\rm e}^ \todaQ)\de z\wedge \de\bar z\, .
\ee
The integrability condition for \eqref{dth} follows from the
three-dimensional continuous Toda equation,
\be
\p_{z} \p_{\bar z}  \todaQ +\p_ \rho^2 \, {\rm e}^ \todaQ = 0\, ,
\label{Toda}
\ee
which is the requirement coming from the Einstein self-duality condition of the metric \cite{MR1423177}.
To see the connection with the more familiar Toda equation from integrable systems, note that
\eqref{Toda} follows from the equation
\begin{equation}\label{cont-Toda}
\p_{z} \p_{\bar z} \cF + {\rm e}^{\p_\rho^2 \cF } = 0\ , \qquad  \todaQ=\p_\rho^2 \cF \, ,
\end{equation}
by differentiating twice with respect to $\rho$. Eq. \eqref{cont-Toda} is recognized as the
continuum limit of the $sl(n)$ Toda system for $n\rightarrow \infty$, and is amenable
to the usual integrable techniques, e.g. its solutions can be  constructed using free fermions
\cite{Jimbo:1983if,Ueno:1984zz,Takasaki:1994xh} .

As stressed already several times, there are infinitely many ways to cast \eqref{dstoda}
into Przanowski form \eqref{metPrz}, corresponding to different choices of integrable
complex structures on $\cM$. However, as explained in \cite{Tod:2006wj}, there is one
canonical choice of complex structure which makes the Killing vector field particular simple
in Przanowski's variables. To specify this complex structure, recall that to
any Killing vector field $\kappa$, one may associate the vector-valued moment map $\vec \mu$ via \cite{MR872143}
\begin{equation}
\label{rmu}
\vec {\muh}_\kappa  = \frac12 ( \vec r_\kappa + \kappa \cdot \vec p )\, ,
\end{equation}
where $\vec p$ is the $SU(2)$ connection and $\vec r$ is a three-vector which
generates the rotation of the quaternionic two-forms under the Lie-derivative along $\kappa$,
\begin{equation}\label{isom-curv}
\cL_\kappa {\vec \omega} + {\vec r}_\kappa \times {\vec \omega}=0\, .
\end{equation}
The moment map $\vec \mu$ determines a
global holomorphic section of $H^0(\cZ,\cO(2))$, represented in the patch
$\cU_i$ by the holomorphic section
\be
\label{defmu}
\muh\ui{i}_\kappa \equiv
{\rm e}^{\Phi\di{i}} \left( \muh^+_\kappa \, \varpi^{-1} -\I \muh^3_\kappa + \muh^-_\kappa \varpi\right)\, .
\ee
In particular, $\muh\di{i}$ encodes the holomorphic action of $\kappa$ on $\cZ$
via $\kappa_\cZ \cdot \hCX\ui{i} = \muh\ui{i}_\kappa$.  Now, the vanishing of $\muh\di{i}$
determines two sections $t_\pm(x^\mu)$, related by the antipodal map, and therefore
one integrable complex structure (and its complex conjugate) canonically associated
to the Killing vector
$\kappa$. Choosing this complex structure \cite{Tod:2006wj}
leads to a Przanowski representation for the metric \eqref{dstoda}
with a "type 1" isometry, i.e. such that $h_1=h_{\bar 1}$ \cite{Przanowski:1991ru}.
The relation between  $h,z^1,z^2$ and $ \todaQ,\rho,z$ is given by
the Lie-B\"acklund transformation
\be
z=z^2\, ,
\qquad
\rho=1/(2h_1)\, ,
\qquad
 \todaQ=h+ 2 \log \rho\, .
\label{Backlund}
\ee
The moment map for the Killing vector $\kappa_1=\I(\partial_{z^1}-\partial_{{\bar z}^1})$
is easily computed to be
\begin{equation}
\vec\mu_{\kappa_1}=-\frac{3}{\Lambda}(0,0,h_1)\, ,
\end{equation}
so that the two sections are $t_+(x^\mu)=0$ and $t_-(x^\mu)=\infty$. Thus,
the complex submanifold $t=0$ plays once again a distinguished r\^ole,
as it produces a Przanowski solution adapted to the Killing field $\kappa$.
Using the identifications \eqref{ressol}, \eqref{rescph} valid for this slice,
the Toda variables $\rho$ and $\todaQ$ can therefore be related to the twistorial data
\be
\rho={\rm e}^\phi\, ,
\qquad
 \todaQ=2\log(\cR/2)\, .
\label{Backlund_tw}
\ee

For a given Przanowski  metric, all isometries are not necessarily of this type however.
Another possibility, but by no means the only one, is if $h_2=h_{\bar 2}$, i.e. if $h$
depends only on $z^2+\bar z^2$: the metric
\eqref{metPrz} now admits a "type 2" Killing field $\kappa_2=\I(\partial_{z^2}-\partial_{{\bar z}^2})$,
with moment map
\begin{equation}
\vec\mu_{\kappa_2}=-\frac{3}{\Lambda}\(\I\,{\rm e}^{h/2},
-\I\,{\rm e}^{h/2},h_2\)\, ,
\end{equation}
corresponding to $t_+(x^\mu)=-\frac12\, {\rm e}^h (h_2 \pm \sqrt{ h_2^2+4 {\rm e}^{2h}})$.
It is not known how to relate such solutions of the master equation \eqref{master} explicitly,
but on general ground there must exist a non-holomorphic change of variables which converts the metric
into Przanowski's form with a type 1 isometry.

If however one knows the twistor space description, such a relation can easily be found.
For example, if the isometry is realized on the twistor space as shifts of the complex coordinate $\alpha$,
one obtains Przanowski's descriptions with ``type 1" or ``type 2" isometries by taking the slices
$\xi=\infty$ or $\xi=0$, respectively. This follows immediately from the definition of the coordinates $z^\alpha$
in \eqref{defcoor} and \eqref{Leg-slice}. Then the knowledge of explicit expressions for these coordinates
allows to find the corresponding non-holomorphic coordinate change.

\subsection{Calderbank-Pedersen meet Toda-Przanowski \label{sec_2iso}}

Self-dual Einstein spaces $\cM$ with
two commuting Killing vectors are described by the Calderbank-Pedersen ansatz \cite{MR1950174},
\be\label{CP-metric}
\begin{split}
{\rm d}s^2=& \frac{3}{\Lambda} \[
\frac{{\cpF}^2-4 \cprho^2({\cpF}_ \cprho^2+{\cpF}_\eta^2)}{4{\cpF}^2}\, \frac{\de \cprho^2+\de\eta^2}{ \cprho^2}\right.
\\
&\left.\qquad
+\frac{\left[\left( {\cpF}-2 \cprho {\cpF}_ \cprho\right) \alpha - 2  \cprho \, {\cpF}_\eta \, \beta\right]^2
+\left[ \left({\cpF}+2 \cprho \, {\cpF}_ \cprho\right) \, \beta - 2  \cprho \, {\cpF}_\eta \,
\alpha\right]^2}
{{\cpF}^2 \left( {\cpF}^2-4 \cprho^2({\cpF}_\cprho^2+{\cpF}_\eta^2) \right)} \] \, ,
\end{split}
\ee
where $x^\mu=\{ \cprho,\eta,\cpphi,\psi\}$ are real coordinates
on $\cM$,
$\alpha=\sqrt{ \cprho}\, \de\cpphi$ and $\beta=(\eta\de\cpphi-\de\psi)/\sqrt{ \cprho}$
are invariant one-forms under the torus action, and
$\cpF( \cprho,\eta)$  is a function
satisfying the Laplace equation on the hyperbolic 2-plane  with a specific eigenvalue,
\be\label{CP-laplace}
 \cprho^2 (\partial_ \cprho^2 +\partial_\eta^2){\cpF}(\eta, \cprho)
=\frac{3}{4}\,{\cpF}(\eta, \cprho)\, .
\ee
As usual, ${\cpF}_ \cprho \equiv \partial_ \cprho {\cpF}$ and
${\cpF}_\eta \equiv \partial_\eta {\cpF}$, etc.

We now single out the Killing vector $\pa_{\psi}$ to cast the metric in Toda's form.
For this purpose, define $G=\sqrt{ \cprho} \, \cpF$. The metric can
be rewritten as \cite{Casteill:2001zk}
\be
\label{dspsi}
\de s^2= -\frac{3}{\Lambda} \[ \frac{P}{G^2} \left( \cprho^2 \de\cpphi^2 + (G_ \cprho^2+G_\eta^2)(\de \cprho^2+\de\eta^2) \right)
+\frac{1}{P G^2} (\de\psi+\Theta)^2\] ,
\ee
where
\be
P=\frac{ \cprho (G_ \cprho^2+G_\eta^2)-G\, G_ \cprho}{ \cprho\, (G_ \cprho^2+G_\eta^2)}\, ,
\qquad
\Theta =\left( \frac{G\, G_\eta}{G_ \cprho^2+G_\eta^2}-\eta\right) \de\cpphi\, .
\ee
Now, introduce new coordinates $z=\haf\(V+\I \cpphi\),\ \rho=G$,  where
$V$ is defined, up to an additive constant, by \cite{MR1950174}
\be
G_ \cprho = - \cprho \, V_\eta\ ,\qquad G_\eta =  \cprho\, V_ \cprho\, .
\ee
The integrability condition for these two equations is $G_{ \cprho \cprho}+G_{\eta\eta}-G_ \cprho / \cprho=0$,
which follows from \eqref{CP-laplace}. Then it is straightforward to check that
\be
 \cprho^2 \de\cpphi^2 + (G_ \cprho^2+G_\eta^2)(\de \cprho^2+\de\eta^2)
= \de \rho^2 + 4 {\rm e}^ \todaQ \de z \de \bar z\, ,
\ee
where $ \todaQ\equiv 2\log \cprho$ can be viewed as a function of $z,\bar z,\rho$
(the additive constant ambiguity in $V$ just amounts to an overall translation of
the solution in the $z$ plane).
Hence \eqref{dspsi} is of the form \eqref{dstoda}, and $\todaQ$
satisfies the Toda equation \eqref{Toda}.
Since $ \todaQ$ is independent of $\psi$, the Przanowski function $h$ defined by \eqref{Backlund}
is independent of $z^2-\bar z^2$. Therefore, the commuting
isometries $\pa_\psi$ and $\pa_\cpphi$ correspond to type 1 and type 2
invariances of the Przanowski function, respectively.

The relation to the Przanowski data and the variables of the twistor formulation
can be found from \eqref{Backlund} and \eqref{Backlund_tw}
\be
\cprho=\frac{{\rm e}^{h/2}}{2h_1}=\frac{\cR}{2},
\qquad
\cpF=\frac{{\rm e}^{-h/4}}{\sqrt{2h_1}}={\rm e}^\phi\sqrt{\frac{2}{\cR}}.
\label{Back_CP}
\ee

\section{Examples}
\label{sec_examples}

\subsection{$S^4$ and $H^4$}

In this subsection we discuss the quaternionic projective space $\IH P^1$ (the four-sphere $S^4$)
and its non-compact analogue, the hyperbolic plane $H^4$. The hyperk\"ahler cones are $\IH^2$ and
$\IH^{1,1}$, respectively. The twistor space above the QK space is obtained  by projectivizing the
HKC with respect to multiplication by complex numbers, so $\cZ=\IC P^3$ or its non-compact version.
This space can be covered by four patches, which we can consider separately to get a global construction of the
four dimensional QK space. The data we need to provide to specify the twistor space $\cZ$ are the
K\"ahler potential $K_\cZ$ and the contact form $\cX$, on each of the patches. Below we give explicit formulae
for one patch, the extension to other patches is relatively straightforward.

As in \cite{Alexandrov:2008ds}, we give two equivalent descriptions, which
both have their own virtues. The first description is the simplest and most commonly used,
but does not obey the scaling properties that are used in the superconformal description
of the hyperk\"ahler cone. Therefore, many of the properties stated in section \ref{sectzero} are not
satisfied. We then show how this can be cured in an improved version that we discuss
at the end of this subsection. The two versions are related by a simple contact transformation.

\subsubsection{First version}

In the first description, we choose local coordinates $u_1,u_2,u_3$ on $\cZ=\IC P^3$ and construct
the Fubini-Study metric with \kahler potential
\be\label{Kpot-patch1}
K = \log \frac12  \left( 1 + |u_1|^2 + \eps |u_2|^2 + \eps |u_3|^2 \right) \, .
\ee
Here, $\epsilon=1$ for $S^4$, and $\epsilon=-1$ for $H^4$. Their twistor spaces are Riemannian or
pseudo-Riemannian respectively. The holomorphic contact form is independent of the signature and
can locally be written as
\be
\cX =\half\( \de u_1 + u_2 \de u_3 - u_3 \de u_2\) \, .
\ee
One can rewrite the contact form to
bring it in Darboux form, by changing variables
\be
\alpha=\half(u_1-u_2u_3)\, ,
\qquad
\xi=u_2\, ,
\qquad
\txi=u_3\, ,
\qquad
\cX=\de\alpha +\xi \de \txi\, .
\end{equation}
Twistor lines can be constructed as
\be
u_1 = t\, ,
\qquad
u_2 = z - \eps \bar w t\, ,
\qquad
u_3 = w + \eps \bar z t \, .
\ee
Notice that they are not of the form \eqref{twist-lines}, as explained in the beginning of this subsection.

The boundary of this patch is when the coordinates $u_1,u_2,u_3\rightarrow \infty$, or in terms of
the twistor lines, at $t=\infty$, that is the southpole on the two-sphere.
Therefore the chosen patch on the twistor space projects to a patch on the twistor sphere which excludes
the southpole. Moreover, for $\eps=-1$, there is an additional restriction.
The K\"ahler potential \eqref{Kpot-patch1} implies that we must restrict ourselves
to the region
\begin{equation}\label{region1}
1+|u_1|^2 > |u_2|^2+|u_3|^2
\ \Rightarrow \
|z|^2+|w|^2<1\, .
\end{equation}
This is precisely the defining equation for a unit four-ball $B^4$. The four-ball is known to be
a stereographic projection of $H^4$, so $z$ and $w$ form a coordinate system on the quaternionic
projective space.

The contact potential can easily be computed from \eqref{c-pot}. We find
\be
{\rm e}^{\Phi} = \frac{t}{4} \left( 1 + \eps (|z|^2+|w|^2) \right)\, .
\ee
Note that it is $t$ dependent. Moreover, substituting the twistor lines into the contact form, we get
the $SU(2)$ connection
\be\label{su2conn-H4}
p^+ = \frac{z \de w - w \de z}{ 1 + \eps (|z|^2+|w|^2)}\, ,
\qquad
p^3 = \I \eps  \frac{z \de \bz + w \de \bar w - \bz \de z - \bar w \de w}{ 1 + \eps (|z|^2+|w|^2)}\, .
\ee

To determine the Przanowski function, we bring $p^+$ to the form \eqref{def-z2}, and follow the
procedure outlined there. The easiest choice is to define the Przanowski coordinates
as\footnote{The coordinate $z^2$ in \eqref{P-coord-patch1} is not well defined for $w=0$. One can cure this by
defining $\tilde z^1=z$ and $\tilde z^2=w/z$. The resulting Przanowski solution in these coordinates
then takes the same form as in \eqref{h-H4patch1}, but now in the tilde-variables. On the overlap, where
$w\neq 0$, we have that $\tilde z^2=-1/z^2$ and $\tilde z^1=-z^1z^2$, so the Przanowski function changes
according to \eqref{equiv-h} with $g(z^2)=-1/z^2$ and $f(z^1,z^2)=-z^1z^2$.}
\begin{equation}\label{P-coord-patch1}
z^1= w\ ,\qquad z^2 = -\frac{z}{w}\, ,
\end{equation}
such that
\begin{equation}\label{h-H4patch1}
h_\cC=2\log \[ \frac{|z^1|^2}{1+\eps |z^1|^2(1+|z^2|^2)}\]\, .
\end{equation}
In terms of slices, this solution can be obtained by taking the special slice $t=0$. Indeed,
it corresponds to the slice $\cC(u)\equiv u_1=0$, and we reproduce \eqref{h-H4patch1}
from \eqref{identh} by noting that $F_\cC=\log \half (z^1)^2$.
Changing $z^1\to {\rm e}^{z^1}$ and dualizing with respect to the type 1 isometry,  \eqref{h-H4patch1}
produces the following solution to the Toda equation \eqref{Toda},
\be\label{Toda-S4}
 \todaQ_\cC=2\log\(\frac{\epsilon}{4}\,\frac{4\rho-1}{1+z\bar z}\)\, .
\ee

Taking other slices yields different solutions of the Przanowski equation. For instance, taking $\cC'(u)\equiv
u_2=0$, one may choose $z'^2=u_1, z'^1=u_3$ and obtain from \eqref{identh}, with $F_{\cC'}=-\log 2$,
\begin{equation}\label{h-H4slice2}
h_{\cC'}=-2\log \( 1 + \epsilon |z'^1|^2 + |z'^2|^2 \)\, .
\end{equation}
Dualizing this into the Toda form leads back to \eqref{Toda-S4} with $\rho\rightarrow -\rho$, which is obviously
still a solution to \eqref{Toda}.
By comparing the twistor lines on the intersection of the two slices, $t=0$ and $t'=\epsilon z/\bar w$,
following the general prescription given in subsection \ref{secnonh}, we find the non-holomorphic
coordinate transformation
\begin{equation}
z'^1=z^1(1+|z^2|^2)\, ,
\qquad
z'^2=-\epsilon\frac{z^1z^2}{\bar z^1}\, ,
\end{equation}
relating the metrics obtained from \eqref{h-H4patch1} and \eqref{h-H4slice2}. This illustrates the general
principle outlined in this paper.

\subsubsection{``Improved" version}

To obtain twistor lines that fall in the class discussed in section \ref{sectzero}, one can
change coordinates to
\be
\xi = \frac{u_2 u_3}{u_1}\, ,
\qquad
\txi = \frac12 \log \frac{u_3}{u_2}\, ,
\qquad
\alpha = \frac12 \log u_1\, .
\label{twistlog}
\ee
This is a contact transformation, in the sense that
\be
\de u_1 + u_2 \de u_3 - u_3 \de u_2 = 2 u_1 \left(
\de\alpha + \xi \de\txi \right)\, .
\ee
In terms of these new Darboux coordinates, the twistor lines now read
\be
\xi = (z - \eps \bar w t)(w + \eps \bar z t)/t\, ,
\qquad
\txi = \frac12 \log \frac{w + \eps \bar z t}{z - \eps \bar w t}\, ,
\qquad
\alpha = \frac12 \log t\, ,
\label{twistimprove}
\ee
which do fall in the class of \eqref{behap}. These expressions for the twistor lines are valid in the patch
away from the cut from $t_+=\eps z/\bar w$ to $t_-=-\eps w/\bar z$.

This description follows from the following construction of the twistor space.
We cover the twistor sphere by three patches: the patches $\cU_{t_\pm}$ surround the points $t_\pm$
and the rest is covered by $\cU_0$.
The two non-trivial transition functions are given by
\be
\Hij{0t_\pm}=\pm\half\, \xi\log\xi\, ,
\ee
and there is a non-vanishing anomalous dimension $\ac=\half$.
Then the general equations \eqref{txiqline} and \eqref{contpotconst}
reproduce\footnote{For that one should use the trick explained in section 3.4 of \cite{Alexandrov:2008ds}
which allows to close the integration contour despite the presence of the logarithmic cuts. This amounts
to integrating the function $\Hij{0t_+}$ around the figure-eight contour surrounding $t_\pm$.}
the twistor lines \eqref{twistimprove} if one identifies
\be
\cR=|zw|\ ,
\qquad
A=\eps(|z|^2-|w|^2)\, ,
\qquad
B=\arg\frac{w}{z}\, ,
\qquad
B_\alpha=\arg(zw)\, ,
\ee
and rotates $t$ by the phase of $zw$: $t\to t\sqrt{\frac{\bz\bw}{zw}}$.

The \kahler potential, after a \kahler transformation, becomes
\be
K= \log\left[ \cosh(\alpha+\bar\alpha)+\epsilon |\xi| \cosh(\txi+\bar\txi)\right]\, ,
\ee
and the contact potential is now $t$ independent,
\be
{\rm e}^{\Phi'} = {\rm e}^{\Phi}/u_1 =  \frac14 \left( 1 + \eps (|z|^2+|w|^2) \right)\, .
\label{contpotlog}
\ee
Choosing the slice $t=0$, the procedure of section \ref{sect0} leads to
\be
z^1=\half\,\log(zw),
\qquad
z^2=\half\, \log\frac{w}{z}
\ee
and
\be
h=-2\log\half\({\rm e}^{-(z^1+\bz^1)}+2\eps \cosh(z^2+\bz^2)\).
\ee

\subsection{The universal hypermultiplet}\label{sect-UHM}

In this section, we illustrate our general results in the case of the perturbative
hypermultiplet moduli space in type IIA string theory  compactified
on a rigid Calabi-Yau three-fold $X$.
For  brevity and following a common abuse of language, we refer to this space as
the ``universal hypermultiplet" space $\cM$.
This is a  (non-complete) \qk deformation of the quaternion-K\"ahler symmetric
space $SU(2,1)/U(2)$, a non-compact version of $\IC P^2=SU(3)/U(2)$. The deformation
is physically interpreted as a one-loop correction to the moduli space metric. It
has a curvature singularity which is presumed to be resolved by instanton corrections
(perturbative corrections vanish beyond one-loop).

The  metric on $\cM$ can
be written in terms of four real coordinates, denoted
by $\{r,\zeta,\tzeta,\sigma\}$,\footnote{They are related to the coordinates used in \cite{Alexandrov:2006hx}
as $\chi=-\frac{ \zeta}{2}$, $\varphi = {\tilde \zeta}$ and our $\sigma$ corresponds to $4\sigma +2\chi\varphi$ there.}
as follows
\be
\de s_{{\rm UHM}}^2=-\frac{3}{2\Lambda r^2}\[\frac{r+2c}{r+c}\,\de r^2+
(r+2c) \frac{|\de{\tilde \zeta} +\frac{\tau}{2}\,\de \zeta |^2}{\Im\tau}+\frac{r+c}{16(r+2c)}
\( \de\sigma+{\tilde\zeta}\de\zeta -\zeta \de{\tilde\zeta}\)^2\]\, .
\label{UHM}
\ee
Here $r={\rm e}^\phi$ represents the four-dimensional string coupling constant
$1/g_4^2$, $\zeta/2,\tzeta$ are the
periods of the Ramon-Ramond three-form on a basis of $H^3(\cX,\IZ)$ and $\sigma$ is the Neveu-Schwarz axion.
The parameter $\tau$ is the modulus of the Jacobian of $\cX$ (see \cite{Bao:2009fg}
for more details); it can be reabsorbed into a linear coordinate transformation of $(\zeta/2,\tzeta)$.
Although it does affect the global structure of $\cM$, for the local
considerations in this work there is no loss of generality in setting $\tau=\I$.
The parameter $c=-\frac{\chi_X}{192\pi}$ is  determined by the one-loop correction to the
tree-level metric,  with $\chi_X$ the Euler number of the Calabi-Yau manifold $X$ \cite{Antoniadis:2003sw}.
For $c=0$, or in the classical limit $r\to\infty$, \eqref{UHM} reduces to the  metric on $SU(2,1)/U(2)$
with isometry group $SU(2,1)$. For $c\neq 0$, the isometry group is broken to the 3-dimensional Heisenberg group
acting as shifts in $\sigma$, $\zeta$ and $\tzeta$,
\be
r\to r,
\qquad
\zeta\to\zeta+\gamma,
\qquad
\tzeta\to\tzeta+\beta,
\qquad
\sigma \to \sigma -4\alpha-\beta\gamma+\gamma\tzeta-\beta\zeta \, ,
\label{HMisom}
\ee
where $\alpha, \beta$ and $\gamma$ are real parameters.
Despite appearances, the \qk metric \eqref{UHM} is regular at $r=0$ and $r=-c$ \cite{Alexandrov:2009qq}.
The curvature singularity at $r=-2c$ is expected to be resolved by instanton corrections.

\subsubsection{Calderbank-Pedersen, Toda and Przanowski}

Since the Killing vectors $\kappa_1=\pa_\sigma$ and $\kappa_2=\pa_{\tzeta}-\zeta \pa_\sigma$ commute, the  metric
can be brought into the Calderbank-Pedersen form  of Section \ref{sec_2iso}.
This is achieved by changing coordinates to \cite{Antoniadis:2003sw,Davidse:2005ef}
\be
\label{coord-CP-UHM}
 \cprho=\sqrt{r+c}\ ,\qquad \eta=\frac{\zeta}{4}\ ,\qquad
\cpphi=\tzeta\ ,\qquad \psi=\frac18(\sigma+\zeta\tzeta)\ .
\ee
The Calderbank-Petersen potential $\cpF$ and Toda potential $\todaQ$ are then given by
\be\label{CP-UHM}
\cpF =  \cprho^{3/2} - c\,   \cprho^{-1/2}\, ,
\qquad
\todaQ=\log(\rho+c)\, ,
\ee
where we have used the fact that $G= \cprho^2-c,\ V=-2\eta$
which are easily seen to satisfy \eqref{CP-laplace}, \eqref{Toda}.

From the general discussion in Section \ref{sec_2iso}, or referring back to
\cite{Alexandrov:2006hx}, we can therefore cast the metric \eqref{UHM}
in Przanowski's form  \eqref{master} by setting
\be
z^1=-\left( r+c\log (r+c)-\frac{1}{8}\,\zeta^2 \right) - \frac{\I}{4}(\sigma+\zeta\tzeta) \, ,
\qquad
z^2=-\frac{1}{4}(\zeta-2\I\tzeta)\, ,
\label{change}
\ee
\begin{equation}
\label{h-one-loop}
h=-\log\frac{r^2}{r+c}\, .
\end{equation}
In this formula, $r$ is the function of $(z^1,z^2)$ and is determined by the following equation
\begin{equation}
r+c\log (r+c) =  \frac{1}{2}(z^2+\bz^2)^2-\half(z^1 +\bz^1)\, .
\label{rstand}
\end{equation}
By computing the derivatives of $h$ implicitly, it is easy to verify that
\eqref{h-one-loop} does solve the master equation \eqref{master}.
For $c=0$, $h$ may be expressed explicitly in terms of $(z^1,z^2)$,
\be
c=0:\qquad h=-\log \[ \half (z^2+\bar z^2)^2-\half (z^1+\bar z^1)\]\, ,
\ee

\subsubsection{Twistor space}

On the other hand, the twistor description for \eqref{UHM} is based on the covering of
the Riemann sphere by three patches \cite{Neitzke:2007ke,Alexandrov:2008nk} :
$\cU_\pm$ are patches around the north and the south pole, whereas the patch $\cU_0$ covers
the rest.\footnote{In principle, the patches $\cU_\pm$ are sufficient  in order to cover
the twistor space of the perturbative universal hypermultiplet. Nevertheless, it is convenient
to introduce the additional patch $\cU_0$ for reasons explained in \cite{Alexandrov:2008nk}.}
Specifying the general results of \cite{Neitzke:2007ke,Alexandrov:2008nk} to the
prepotential $F(X)=\frac{\I}{4}\,X^2$, one obtains that
the transition functions between these patches read
\be
\label{symp-cmap}
\Hij{0+}= -\frac{1}{8}\, \xi^2\, ,
\qquad
\Hij{0-}=\frac{1}{8}\,\xi^2\, .
\ee
In addition, $\alpha$ has anomalous dimension, due  the one-loop correction,
\be
\tc=0\, ,
\qquad
\ac = -2c \, .
\ee
They lead to the following Darboux coordinates
in the patch $\cU_0$\footnote{To simplify
the notations, we will omit the index $\scriptstyle [0]$ for the twistor lines in this patch.}
\begin{eqnarray}\label{classical-twistors1}
\xi&=&\zeta+2{\sqrt{r+c}}\,(t^{-1}-t)\, ,
\qquad
{\tilde \xi}= \frac{\I}{2}\,{\tilde \zeta}+ \frac12{\sqrt{r+c}}\,(t^{-1}+t)\, ,
\nonumber\\
\alpha+\hf\,\xi\txi&=&-\frac{\I}{4}\,\sigma-\frac{1}{4}\,\sqrt{r+c}\((\zeta-2\I\tzeta)t^{-1}+(\zeta+2\I\tzeta)t\)
-2c\log t\, .
\end{eqnarray}

Following the same procedure as in the previous subsection, one can compute the contact potential and
the $SU(2)$ connection one-forms. The contact potential is real,
constant along the fiber, and coincides with the four-dimensional dilaton
\begin{equation}
{\rm e}^{\Phi}=r\, ,
\label{phiuhm}
\end{equation}
so that the K\"ahler potential is given by
\be
\label{KUH}
K = \log\frac{2r ( 1+t\bt)}{|t|} \, .
\ee
It is impossible to write it explicitly as a function of Darboux coordinates and their
complex conjugate. This can be done however for $c=0$, in which case it reads as \cite{Gunaydin:2007qq}
\be\label{K-UHM}
K = \frac12\, \log\left[\left( \frac{1}{16}\left(\xi-\bar\xi\right)^2 - \left(\txi+\bar\txi\right)^2 \right)^2
- \left( \alpha +\bar\alpha + \frac12(\xi+\bar\xi)(\txi+ \bar\txi)\right)^2 \right]\, .
\ee
One can explicitly check that the Monge-Amp\`ere equation \eqref{MAeq} is satisfied, as well as the other identities
on the twistor space \eqref{defomm}.
The $SU(2)$ connection is easily found by expanding the contact form,
\be
\label{p-uhm2}
p^+ = -\frac{\sqrt{r+c}}{4r} ( \de\zeta - 2\I \de \tzeta)\, ,
\qquad
p^3=\frac{1}{8r}
(\de\sigma+\tzeta \de\zeta - \zeta \de\tzeta)\, .
\ee


One may check that the Przanowski description presented above
follows from this twistor construction by restricting to the $t=0$ slice.
In particular, the function \eqref{h-one-loop} follows from \eqref{relhPhi}
and the coordinates \eqref{change} coincide with \eqref{defcoor} up to a constant shift in $z^1$.
It is also easy to verify that the relations \eqref{Backlund_tw}, \eqref{Back_CP} to the Toda
and the Calderbank-Pedersen variables, respectively, are indeed satisfied.

\subsubsection{Alternative Przanowski description}

The Przanowski function \eqref{h-one-loop} has a type 1 Killing vector $\kappa_1=\pa_\sigma$, and is
canonically associated to $\kappa_1$ along the lines of Section \ref{secTod}. The other Killing
vector $\kappa_2=\pa_{\tzeta}-\zeta \pa_\sigma$ is realized as a type 2 isometry. There is however
a different Przanowski representation of the metric \eqref{UHM} canonically associated to
$\kappa_2$, i.e. where $\kappa_2$ becomes a type 1 isometry. Since the moment map associated
to $\kappa_2$ is proportional to $\xi$, this representation may be reached by choosing the
complex submanifold $\xi=0$. More generally, the complex submanifold $\xi=x$ for any fixed
complex number $x$ produces a Przanowski representation of the metric \eqref{UHM},
which interpolates between the two afore-mentioned representations. For simplicity, we restrict to
the case with $c=0$.

The slice $\xi=x$ is an example of the $\hat\xi$-parametrization
discussed in Section \ref{subsubsec_xi}, and leads to the Przanowski description
provided by \eqref{Leg-slice} and \eqref{hW} with $W=x\txi$.
This gives
\be
\label{tz12}
z'^1 = \txi\, ,
\qquad
z'^2 = \alpha+x \txi \, .
\ee
and produces the following solution of the master equation
\be
h'(z'^1,z'^2) = - \log\left[ \left( (z'^1+\bz'^1)^2 - \frac{(x-\bar x)^2}{16}\right)^2
- \left(z'^2+\bz'^2-\frac{x-\bar x}{2}\,(z'^1-\bz'^1)\right)^2 \right].
\label{hnew}
\ee
The coordinates $z'^\alpha$ are related to the real coordinates $x^\mu$ in \eqref{UHM} by
solving the condition $\xi=0$ for $t(x^\mu)$ and plugging into \eqref{tz12},
\bea
z'^1 & = &\frac{\I}{2}\,\tzeta + \frac{1}{4} \sqrt{16 r + (\zeta-x)^2}\, ,\qquad
\\
z'^2+ x z'^1 & = &-\frac{\I}{4}\(\sigma+\zeta\tzeta\) - \frac{1}{8} (\zeta+x)\sqrt{16 r + (\zeta-x)^2}\, .
\label{zt12}
\eea
In particular, as advertised previously, the Killing vector $\kappa_2$ is now type 1.
The fact that $h'$ is independent of $x+\bar x$ follows from the fact that $\Re(x)$ can be
reabsorbed into a shift of $\zeta$. In particular, for $x=0$, the coordinates
\eqref{zt12} are related to \eqref{change} by the  non-holomorphic change of variables
\be
\begin{split}
z'^1 &= \frac12 \left( \bz^2 - z^2\right) +\frac14 \sqrt{16 (z^1+\bz^1)+12(z^2+\bz^2)^2}\, ,
\\
z'^2 &=  (\bz^1- z^1) - \frac{2(z^2+\bz^2)}{8}
\sqrt{16 (z^1+\bz^1)+12(z^2+\bz^2)^2}\, ,
   \end{split}
\ee
which illustrates the general situation described in Section \ref{secnonh}.

\section{Linear perturbations \label{seclin}}

In this section, we present one of the main results of the paper.
Namely, we analyze linear perturbations of self-dual Einstein manifolds with an isometry. The general
theory of quaternionic perturbations was developed in \cite{Alexandrov:2008nk}, which we specify here to
the case of four dimensions.
We show that the linearized master equation \eqref{eqdPrz} is equivalent, after
suitable rescaling of $\delta h$, to the vanishing of the conformal
Laplace-Beltrami operator on $\cM$. We further express the variations $\delta h$ through
the Penrose type integral of a set of holomorphic functions representing a section of $H^1(\cZ,\cO(2))$,
which governs the perturbations of the twistor space $\cZ$.

\subsection{Linearized master equation and conformal massless field}

To begin, we observe that in the background of self-dual Einstein manifold $\cM$
with a type 1 isometry, the linearized master equation \eqref{eqdPrz} controlling
perturbations of the \qk metric on $\cM$ has a simple
relation to the Laplace-Beltrami operator on $\cM$. Indeed,
noting that the inverse Przanowski metric is given, in $\p_1,\p_2$ basis, by
\be
g^{\alpha\bar\beta} = -\frac{\Lambda}{3} \frac{{\rm e}^{-h}}{|h_1|^2}
\begin{pmatrix} h_{2\bar 2} + 2 {\rm e}^h & - h_{2\bar 1}
\\
-h_{1\bar 2} & h_{1\bar 1} \end{pmatrix}\, ,
\ee
 the Laplace-Beltrami operator on $\cM$ reads
\be
\label{laplace}
\begin{split}
\Delta &=
\frac{1}{\sqrt{g}}\, \p_{\mu}\(\sqrt{g}g^{\mu\nu}\p_{\nu}\)
\\
&= - \frac{ 2\Lambda\, {\rm e}^{-h}}{3 |h_1|^2}
\left[ h_{1\bar 1} \p_{2 \bar 2} +  h_{2\bar 2} \p_{1 \bar 1}
- h_{1\bar 2} \p_{2 \bar 1} - h_{2\bar 1} \p_{1 \bar 2}
+ {\rm e}^h \left( h_1 \p_{\bar 1} + h_{\bar 1} \p_{1} + 2 \p_{1\bar 1} \right)  \right]\, .
\end{split}
\ee
It is then a simple matter to check that, when\footnote{This identity
is the specialization to the type-1 isometric case
of a more general statement which holds for any solution of the master equation,
\be
\dPrz_h \, (\delta h)=  {\rm e}^h \, |h_1|^4 \,
\[  -\frac{3}{2\Lambda} \tilde \Delta + 1 \]\frac{\delta h}{|h_1|^2},
\label{geneq}
\ee
where the operator
\be
\tilde \Delta=\Delta
+2\cV_\mu \, g^{\mu\nu}
\left(\p_{\nu}+\frac12 \(\p_{\nu}\log|h_1|^2\) \right),
\qquad
\cV_\alpha=\p_{\alpha}\log\frac{h_1}{h_{\bar 1}},
\label{genop}
\ee
reduces to $\Delta$ when $h_1=h_{\bar 1}$. However, the geometric interpretation of this
more general operator is unclear in the absence of type-1 isometries.
} $h_1=h_{\bar 1}$,
\be
\dPrz_h \, (\delta h)=  {\rm e}^h \, |h_1|^4 \,
\[ -\frac{3}{2\Lambda} \Delta + 1  \]\frac{\delta h}{|h_1|^2}\, .
\label{geneqiso}
\ee
Thus, in this case, the metric perturbations preserving the QK property (but not
the isometries) are, up to rescalings by $|h_1|^2$, eigenmodes of the Laplace-Beltrami operator,
with a specific eigenvalue equal to $R/6=2\Lambda/3$. This value corresponds to
a conformally coupled massless scalar field, such that  \eqref{geneqiso} is invariant
under conformal rescalings of the background metric \cite{Neitzke:2007ke}.

\subsection{Penrose contour integrals}

As explained e.g. in \cite{Penrose:1972ia,Eastwood:1981jy,Neitzke:2007ke}, eigenmodes
of the  Laplace-Beltrami operator\footnote{In dimension
$4n>4$, Eq. \eqref{Pen0} in fact produces solutions of the stronger condition
$(\epsilon^{A'B'} \nabla_{A A'} \nabla_{B B'} -\nu \eps_{AB}) \psi=0$, which implies
$[\Delta - R/(2(n+2))]\psi=0$   \cite{Neitzke:2007ke}.} on a \qk manifold
with this particular eigenvalue are known to be generated by a Penrose-type contour
integral formula \cite{Neitzke:2007ke} from a holomorphic section of $H^1(\cS)$,
homogeneous of degree 2. This section may be described locally by holomorphic functions
$\hat \Psi\ui{ij}$, subject to the co-cycle relation $\hat \Psi\ui{ij}+\hat \Psi\ui{jk}=\hat \Psi\ui{ik}$. The
solution $\psi$ is then obtained by the contour integral formula
\be
\label{Pen0}
\psi=  \frac{1}{2\pi\I} \sum_j \oint_{C_j}
 \pi_{A'} D\pi^{A'} \hat \Psi\ui{ij}(\pi,x^\mu)\ ,\qquad \[ -\frac{3}{2\Lambda} \Delta + 1  \]\psi=0\, ,
\ee
where $C_j$ are closed contours around each of the patches $\cU_j$
and $\pi_{A'} D\pi^{A'}$ is the Liouville form on $\cS$ (note that
the r.h.s is independent of the value of $i$ due to the co-cycle relations.). Using \eqref{pisec},
it is easily seen to be equal to
\be
\pi_{A'} D\pi^{A'} = {\rm e}^{2\lambda^{[j]}}\, \cX^{[j]} = 2 \, {\rm e}^{2\lambda^{[j]}+\Phi^{[j]}} \frac{Dt}{t}\, .
\ee
Moreover, since $\hat \Psi\ui{ij}(\pi,x^\mu)$ is holomorphic and homogeneous of degree 2, it
can be expressed as
$\hat \Psi\ui{ij}(\pi,x^\mu)={\rm e}^{-2\lambda\ui{j}} \Psi\ui{ij}(\xi,\txi,\alpha)$, where
$\Psi\ui{ij}(\xi,\txi,\alpha)$ can be viewed as
an element of  $H^1(\cZ,\cO(-2)))$. Putting these facts together, the Penrose integral
\eqref{Pen0}  may be rewritten as
\be
\label{Penr1}
\psi=  2 \sum_j \oint_{C_j} \frac{{\rm d}t}{2\pi\I t}\, {\rm e}^{\Phi\ui{j}} \,\Psi\ui{ij}(\xi,\txi,\alpha) \, ,
\ee
where $\Phi,\xi,\txi,\alpha$ are viewed as meromorphic functions of $t$, for a fixed point $x^\mu$
on the base $\cM$.  In the case of an Einstein self-dual metric with a type 1 isometry, this may be
rewritten using \eqref{rescph} as
\be
\label{Pen1}
\psi=  h_1^{-1} \sum_j \oint_{C_j} \ \frac{{\rm d}t}{2\pi\I t}  \Psi\ui{ij}(\xi,\txi,\alpha) \, .
\ee

On the other hand, we know from  \cite{lebrun1994srp,Alexandrov:2008nk} that
deformations of the twistor space $\cZ$ are governed by an element of
$H^1(\cZ,\cO(2))$. This can be represented locally by holomorphic functions
$\Hpij{ij}(\xi,\txi,\alpha)$, defined on the overlap of two patches $\cU_i \cap \cU_j$,
which determine the perturbation of the
holomorphic transition function $\Hij{ij}$ from section 2.2. The perturbed
contact potential is then obtained by expanding \eqref{solcontpot}, \eqref{contpotconst}
to linear order in perturbation. In the case when the background has one isometry,
we can use the results   \eqref{relhPhi}, \eqref{defcoor}, \eqref{ressol} to obtain
the perturbation of the Przanowski function,
\be
\delta h =-2\delta \phi-h_\alpha \delta z^\alpha-h_{\balp}\delta\bz^{\balp}
= -{\rm e}^{-\phi} \[2\delta {\rm e}^\phi +\(\Re\delta\ai{+}_0+\xii{+}_0\delta\txii{+}_0\)\].
\label{varhphi}
\ee
Keeping the real coordinates $\cR, A,B,B_\alpha$ unperturbed and using the integral equations \eqref{txiqline},
one finds
\be
\delta h =-h_1\sum_j \int_{C_j}\frac{\d t}{2\pi \I t} \( \Hpij{+j}
+\delta\txii{j}\p_{\txii{j}}\Hij{+j}-\(\xii{+}-\xi_{(0)}\)\delta\tT^{[+j]}\) ,
\label{inter_form}
\ee
where $\xi_{(0)}=A+\cR\(t^{-1}-t\)$. This can be further simplified by substituting
integral representations for $\delta\txi$ and $\xi-\xi_{(0)}$.
Then the second and third terms differ essentially only by the order of integrations.
They exactly cancel each other since the contribution of the pole at $t=t'$ coming from changing
the order of integration is canceled by the term accounting for the difference between
$\xii{+}$ in \eqref{inter_form} and $\xii{j}$ to be used in this cancelation.
Thus, one finally obtains
a very simple result\footnote{It is not difficult to generalize \eqref{delh} to include the variations
of anomalous dimensions. In this case the result takes the form
\be
\delta h=-h_1\[\oint\frac{\d t}{2\pi \I t} \( \Hp  +\delta \tc \log t\, \p_{\txi}H \)
+(2+h-\log h_1^2)\delta\ac+\frac{h_2+h_{\bar 2}}{2h_1}\(h-\log h_1^2\)\delta\tc\].
\ee
Although the anomalous dimension terms are not integrals of a holomorphic function, they
are annihilated by the conformal Laplacian.}
\be
{\delta h}=-h_1\sum_j \oint_{C_j} \frac{\d t}{2\pi \I t}\, \Hpij{ij}.
\label{delh}
\ee

Thus, comparing \eqref{delh} with \eqref{Pen1} and using \eqref{geneqiso}, we may identify
\be\label{idpsi}
\psi = \frac{\delta h}{|h_1|^2}\ ,\qquad  \Psi\ui{ij} = - \Hpij{ij}.
\ee
This identification between $H^1(\cZ,\cO(2))$ and $H^1(\cZ,\cO(-2))$ of course relies on
the particular trivialization $\lambda=0$ chosen to relate $\hat \Psi$ and $\Psi$.
Using \eqref{idpsi},
we can in principle lift any solution of the linearized master equation \eqref{eqdPrz}
to a section of $H^1(\cZ,\cO(2))$, and therefore infer the contact structure on the
deformed twistor space.

\subsection{Instanton corrections to the universal hypermultiplet}

To illustrate the above results, we now make contact with the
analysis of instanton corrections to the universal hypermultiplet in \cite{Alexandrov:2006hx}.
To begin however, note that for the metric \eqref{UHM},
one may directly prove that the contour integral \eqref{Pen1}
produces eigenmodes of the conformal Laplacian, using the facts that $h_1=1/(2r)$ and
\be
\left[-\frac{3}{2\Lambda} \Delta + 1 - \pa_t \left( \frac{r^2}{4(r+c)(r+2c)} ( t^2 \pa_t + t) + 4 \I t \frac{r^2}{r+c}
 \pa_\sigma) \right) \right] \( \frac{r}{t}\, \Psi\) = 0,
\label{lapl_hol}
\ee
for any holomorphic function $\Psi(\xi,\txi,\alpha)$. If $C_j$ are closed contours,
one may integrate by parts and check that the conformal Laplace equation \eqref{Pen0}
is satisfied. One may also consider open contours, provided one makes sure that
all boundary contributions coming from the integration by parts and the action of the Laplace
operator on the limits of integration cancel each other.
It can be demonstrated
by a rather lengthy computation that this is the case when the end points of the contour lie
on any complex submanifold of $\cZ$.\footnote{This fact can be understood heuristically as follows.
Open contour integrals are expected to arise from closed
contour integrals, associated to standard coverings of $\cZ$ by open patches, upon shrinking the
integration contour around  branch cuts in the transition functions. Since the transition
functions are holomorphic sections of some line bundle on $\cZ$, the end points of the cuts, and
therefore of the open contours, must be given by holomorphic sections $t(x^\mu)$
of $\cZ$.}
Such open contours will play an important role in
what follows.

As a first example, we choose a set of holomorphic functions independent of $\alpha$, of the form
\begin{equation}
\label{d2h}
\Psi_{p,q}(\xi,\tilde\xi)={\rm e}^{\I q\xi - 2 p\txi}\, ,
\end{equation}
for arbitrary "charges" $p,q$. We integrate it over the contour $C$ going from $t=0$ to $t=\infty$
along the direction $(p-\frac{\I}{2}\, q)t^{-1}\in \IR^+$
\be
\delta h_{p,q} =h_1 \, \int_0^\infty \frac{\de t}{2\pi\I t}\, {\rm e}^{\I q\xi - 2 p\txi} \, .
\label{intd2h}
\ee
By using \eqref{lapl_hol}, this will produce an eigenmode of the conformal Laplacian;
the boundary terms vanish since the exponential is exponentially suppressed at $t=0,\infty$ along $C$.
The integral \eqref{intd2h} can now easily be done and produces a modified Bessel function
with index 0,
\be\label{d2}
\delta h_{p,q} = \frac{{\rm e}^{i(q\zeta-p\tzeta)}}{4\pi \I r}\, K_0\left( 2 \sqrt{(r+c)(4q^2+p^2)} \right) .
\ee
In \cite{Alexandrov:2006hx}, this deformation was found to describe a single D-brane
instanton\footnote{The inclusion of multi-coverings turns $\Psi$ into a dilogarithm sum \cite{Alexandrov:2008gh}.}
correction, coming from a wrapped D2-brane on the cycle $(p,q)\in H_3(X,\IZ)$. The same
holomorphic function as in \eqref{d2h}
also governs the complex contact transformations on $\cZ$, and in particular determines the
D2-brane corrections to
contact potential \eqref{contpotconst} in terms of a modified Bessel function with index 1 \cite{Alexandrov:2008gh}.
It is straightforward to check that this is indeed
consistent with the formula \eqref{relhPhi} relating the contact potential and the
Przanowski function in the special slice $t=0$. In checking this one must
take into account the deformation of the coordinates $z^\alpha$ as in \eqref{varhphi}.

As a second example, we consider, for $k>0$ and $\nu\in\IC$,
\be
\Psi_{k,\pm}^{(\varsigmapar)} = ({\textstyle\frac{1}{4}}\,\xi \pm \txi )^\varsigmapar \,
{\rm e}^{\mp 4 k(\alpha+\hf\,\xi\txi) -k\(\frac{1}{4}\xi^2-4\txi^2\)}\, ,
\label{NS5fun}
\ee
and choose a contour $C$ connecting $t=\infty\, (t=0)$
to the point $t_{\pm}(x^\mu)$ on the complex submanifold $\xi\pm 4\txi=0$, namely.
\be
t_{\pm}=-\[\frac{4\sqrt{r+c}}{\pm\zeta+ 2\I\tzeta}\]^{\pm 1}.
\ee
Changing the integration variable as $t=t_\pm (s+1)^{\pm 1}$, one finds
\be
\delta h_{k,\pm}^{(\varsigmapar)} =C\,
\frac{({\scriptstyle\frac{1}{4}}\,\zeta\pm {\scriptstyle\frac{\I}{2}}\,\tzeta )^{\varsigmapar-8ck}}{r\sqrt{r+c}}\,
W_{4ck-\varsigmapar-\hf,4ck}\(8k(r+c)\)\, {\rm e}^{-k\({\scriptscriptstyle\frac{1}{4}}\,\zeta^2+\tzeta^2\) \pm i k\sigma  }\, ,
\label{NS5dh}
\ee
where $C$ is an irrelevant constant and the Whittaker function is defined as the integral
\be
W_{\ell,m}(z)=\frac{{\rm e}^{-z/2}z^{m+\hf}}{\Gamma(\half -\ell+m)}\,\int_0^\infty s^{m-\ell-1/2}(s+1)^{m+\ell-1/2}
{\rm e}^{-zs}\,\de s\ .
\ee
This result reproduces the family of solutions discussed in  Eq. B.19
in  \cite{Alexandrov:2006hx} for $\kappa=8ck-\varsigmapar$.
In the weak coupling limit $r\to\infty$, $\delta h_{k,\pm}^{(\varsigmapar)}$ is exponentially suppressed
\be
\delta h_{k,\pm}^{(\varsigmapar)}\sim
\frac{({\scriptstyle\frac{1}{4}}\,\zeta\pm {\scriptstyle\frac{\I}{2}}\,\tzeta )^{\varsigmapar-8ck}}{r (r+c)^{1+\varsigmapar-4ck}}\,
{\rm e}^{-4kr -k\({\scriptscriptstyle\frac{1}{4}}\,\zeta^2+\tzeta^2\) \pm i k\sigma  }\ ,
\ee
and the argument of the exponential agrees with the NS5-brane classical action.
Note that \eqref{NS5dh} has branch cuts in the $(\zeta,\tzeta)$
plane, a feature which may be undesirable for NS5-brane instanton corrections.
However, for the special value $\varsigmapar=8ck$ in  \eqref{NS5dh}, the branch cut disappears and the instanton correction
becomes particularly simple. In this case it can be written also through
the incomplete Gamma function
\be
\delta h_{k,\pm}^{(8ck)}  =C r^{-1}(r+c)^{4ck} \,  \Gamma\left(-8ck, 8k(r+c)\right)\,
{\rm e}^{4kr -k\({\scriptscriptstyle\frac{1}{4}}\,\zeta^2+\tzeta^2\) \pm i k\sigma  }\ ,
\label{symsimsol}
\ee
where $\Gamma(s,x)=\int_x^\infty t^{s-1}{\rm e}^{-t} {\rm d}t$ is the incomplete Gamma function.
This  reproduces the solution (4.6) in \cite{Alexandrov:2006hx}. For $c=0$ it also agrees with
Eq. (4.72) of \cite{Bao:2009fg}, upon setting $s=\ell_1=\ell_2$ in this reference.

More generally, one would like to determine the holomorphic functions corresponding
to a complete basis of solutions of \eqref{eqdPrz}, regular in the $\zeta$ plane, and exponentially
decaying at $r=\infty$ and $\zeta=\infty$.
A particularly convenient basis can be found by separation of variables, and
is given by
\be
\delta h_{k,n,\pm} = r^{-1}(r+c)^{4ck}\, H_n(\zeta\sqrt{k})\, U\left( 1+n+8ck, 1+8ck,8k(r+c)\right)\,
{\rm e}^{-4 k r -k\frac{\zeta^2}{2}\pm i k(\sigma+\zeta\tzeta)  }\, ,
\label{hermsol}
\ee
with $n\in\IN$,  $H_n$ are the Hermite polynomials
and $U$  is the confluent hypergeometric function of the second kind.
For $n=0$, this reduces to Eq. (4.9) in  \cite{Alexandrov:2006hx},
while for $c=0$, it reduces to the non-Abelian Fourier eigenmodes in  Eq. (1.18) of \cite{Bao:2009fg},
again setting $s=0,n=0$ in this reference. Some results can be
obtained to find a representation of the holomorphic section
$\Psi_{k,n}$ governing \eqref{hermsol} for $n=0$, but the generic problem for arbitrary values of $n$ remains an
interesting open problem for future research.

\section{Discussion}

In this work, we have discussed Einstein self-dual manifolds $\cM$, the four-dimensional avatar
of \qk manifolds, with particular emphasis on their Heavenly description, namely as solutions
of Przanowski's equation \eqref{master}. In particular, we have related this description to
the more standard twistor construction for \qk manifolds, and have shown that the
Przanowski function $h$ was equal to the \kahler potential $K$ on $\cZ$ in a certain \kahler gauge,
restricted to any complex (local) submanifold  $\cC$ of $\cZ$, Eq. \eqref{identh} above.
Different choices of $\cC$ lead to diffeomorphic hermitian metrics on $\cM$, where different
(local, integrable) complex structures are manifest. Varying $\cC$ leads to ``pure gauge" solutions
of the linearized master equation \eqref{eqdPrz}, which lie outside the class of infinitesimal
holomorphic diffeomorphisms \eqref{diff-equiv}. These ``pure gauge" solutions however do not
seem to be expressible in terms of $h$ and its derivatives in general, see \eqref{delhnonhol} and \eqref{delhK3}
for two illustrative examples.

If $\cM$
admits a a Killing vector, there is a preferred choice of submanifold $\cC$, the zero locus of the $\cO(2)$
valued moment map, which determines a canonical complex structure (up to complex conjugation).
In this complex structure, the Przanowski function has a type 1 symmetry ($h_1=h_{\bar 1}$),
and determines a solution of the continuous Toda equation \eqref{Toda} via \eqref{Backlund}.
This reproduces  Tod's parametrization  \eqref{dstoda} of Einstein self-dual manifolds with one
isometry. If $\cM$ admits two commuting isometries, then we have shown that it could be
represented by a  Przanowski function with both type 1 and type 2 ($h_2=h_{\bar 2}$)
symmetries. Such a function determines a solution of the Laplace equation on the
Poincar\'e upper half-plane \eqref{CP-laplace}, reproducing  the Calderbank-Petersen
parametrization  \eqref{CP-metric} of Einstein self-dual manifolds with two commuting isometries.
To our knowledge, the relation between the Calderbank-Petersen potential $\cpF$
and the Toda potential $\todaQ$ has not appeared previously in the literature.

By a similar reasoning, a Killing tensor of higher rank on $\cM$ would determine
a $2n$-plet of locally integrable complex structures. Indeed, for a rank $n$ Killing tensor,
there is a variant of the moment map construction, which is
now a $\cO(2n)$ global section $\mu$ \cite{MR1848654,dunajski2003tth}.
Any of the  $2n$ zeros of $\mu$
determines a locally integrable complex structure. It would be interesting to understand
how the generalized Killing symmetry constrains  the corresponding Przanowski function.

Having understood the twistorial origin of the Heavenly description, we were able to
cast solutions of the linearized master equation \eqref{eqdPrz} around an Einstein
self-dual metric with one Killing vector into the general formalism for perturbations of \qk
manifolds developed in \cite{Alexandrov:2006hx}. To this aim, we observed that the
linearized master equation around such a background is equivalent to the conformal
Laplace-Beltrami operator on $\cM$, Eq. \eqref{geneqiso}, the zero-modes of which can be
obtained by a Penrose-type contour integral \eqref{Penr1} of an holomorphic section
$\Psi\in H^1(\cZ,\cO(-2))$. By studying perturbations of the twistor lines on the special
complex locus $t=0$, we were able to relate $\Psi$ to the holomorphic section
$\Hp\in H^1(\cZ,\cO(2))$ which governs the perturbations of the complex contact
structure on $\cZ$. This relation in principle enables us to lift any solution of  \eqref{eqdPrz}
to a perturbation of the twistor space $\cZ$.

For perturbations around self-dual Einstein manifolds without isometry, the equivalence
between the linearized master equation and the conformal
Laplace-Beltrami operator no longer holds. Instead, one must replace the
Laplace-Beltrami operator by $\tilde \Delta$, defined in   \eqref{genop}.
It would be interesting to understand $\tilde \Delta$ geometrically,
and provide a Penrose-type integral formula for general solutions of \eqref{geneq}.

The main motivation for this work was to understand the structure of instanton corrections
in hypermultiplet moduli spaces in string theory. In Section 6 we applied our results to the
special case of the "universal hypermultiplet", i.e. the hypermultiplet
moduli space in type IIA string theory compactified on
a rigid Calabi-Yau manifold $X$, a particular example of a self-dual Einstein space with
negative curvature. In particular, we showed that the solutions \eqref{d2} of the linearized
master equation corresponding to D2-brane instantons are consistent with the analysis
in  \cite{Alexandrov:2008gh}. We have also taken some steps in lifting the
solution \eqref{hermsol}, which should physically correspond to NS5-brane solutions, to the twistor
space. We have succeeded for the related ``symmetric gauge" solution \eqref{symsimsol}.
Such a solution (for $c=0$) appears in the Fourier expansion of the Picard Eisenstein
series which was considered in \cite{Bao:2009fg}, though it is not well suited for
a systematic analysis of the non-Abelian Fourier expansion.

Eventually, one would like to be able to construct the exact quantum corrected metric
on the hypermultiplet moduli space in rigid Calabi-Yau compactifications. This could
be achieved by determining the exact contact structure on the twistor space, by providing
a set of complex contact transformations consistent with the cocycle condition, such that
the metric reduces to \eqref{UHM} in the weak coupling limit $r\to\infty$, up to
exponentially suppressed corrections of the form discussed above.
This finite deformation of \eqref{UHM} should tame the divergence of the D-instanton
series \cite{Pioline:2009ia}, and resolve the curvature singularity at $r=-2c$.
When $\tau=\sqrt{-d}$ where $d$ a positive integer, it was argued in  \cite{Bao:2009fg}
that this exact metric would be determined by automorphy under the Picard modular
group $SU(2,1,\IZ[\tau])$. The analysis of \cite{Bao:2009fg} however
remained suggestive only due to the difficulties
of implementing discrete symmetries at the level of twistor space (see \cite{Alexandrov:2009qq}
for recent progress in the case of $SL(2,\IZ)$). Hopefully the
results in this paper will be useful in this direction.

More ambitiously, one would like to determine the exact quantum corrected metric
on the hypermultiplet moduli space for general, non-rigid Calabi-Yau compactifications.
In this respect, we note that the Heavenly parametrization generalizes straightforwardly to
\qk manifolds with dimension $4n>4$, though the master equation \eqref{master} is replaced
by $2n-1$ partial differential equations. It would be interesting to determine the linearized
perturbations corresponding to NS5-brane instantons, and their lift to the twistor space.

\acknowledgments

We are grateful to D. Persson for valuable discussions, and to A. Neitzke and F. Saueressig
for past collaboration on closely related topics. S.V. is grateful to LPTHE
and LPTA for hospitality while this project was completed.

\appendix

\section{Derivation of the Przanowski equation in the $\xi=0$ gauge}

The complex slice $\xi=0$ allows for a simple ``top-down" derivation of the Przanowski equation
from the twistor space constraints as follows.
Eqs. \eqref{constrK} reduce to
\be
K^{\txi\balp} = 2\, {\rm e}^{2K} K_\xi \, ,
\qquad
K^{\xi\balp}\ = -2 {\rm e}^{2K}\, K_{\txi} \, ,
\qquad
K^{\alpha\balp} =  {\rm e}^{2K}\, .
\ee
Using the comatrix formula for $K_{\txi\bar\txi}$ in terms of $K^{i\bar j}$
and the Monge-Amp\'ere equation \eqref{MAeq},  we readily obtain
\be
K^{\xi\bar\xi} =4\, {\rm e}^{2K} \, \left( K_{\txi\bar\txi}+ K_{\txi} K_{\bar\txi} \right)\, .
\ee
Requiring that $K^{i\bar j} K_{\bar j k}=\delta^i_k$ produces 4 linear equations in
$K_{\xi\bar\txi}, K^{\xi\bar\txi},K_{\bar\alpha\xi},K_{\bar\xi}$:
\bea
   {\rm e}^{2 K} K_{\alpha\bar\txi}-2 {\rm e}^{2 K}
   K_{\bar\txi} K_{\xi\bar\txi}+2 {\rm e}^{2 K}
   K_{\bar\xi} K_{\txi\bar\txi}&=&0\, ,
\\
   {\rm e}^{2 K}
   K_{\alpha\bar\alpha}-2 {\rm e}^{2 K} K_{\bar\txi}
   K_{\xi\bar\alpha}+2 {\rm e}^{2 K} K_{\bar\xi}
   K_{\txi\bar\alpha}-1&=&0\, ,
\\
   -2 {\rm e}^{2 K} K_{\txi}
   K_{\alpha\bar\alpha}+4 {\rm e}^{2 K} K_{\xi\bar\alpha}
   (K_{\txi\bar\txi}+K_{\txi}
   K_{\bar\txi})+K_{\txi\bar\alpha}
   K^{\xi\bar\txi}&=&0\, ,
\\
   -2 {\rm e}^{2 K} K_{\txi}
   K_{\alpha\bar\txi}+4 {\rm e}^{2 K} K_{\xi\bar\txi}
   (K_{\txi\bar\txi}+K_{\txi}
   K_{\bar\txi})+K_{\txi\bar\txi}
   K^{\xi\bar\txi}&=&0\, .
\eea
A non-zero solution exists when the discriminant vanishes,
\be
\label{przxi0}
 {\rm e}^{2 K}
   K_{\alpha\bar\txi}
   K_{\txi\bar\alpha}+K_{\txi\bar\txi}+
   K_{\txi} K_{\bar\txi}- {\rm e}^{2 K}
   K_{\alpha\bar\alpha} K_{\txi\bar\txi} = 0\, .
\ee
This reproduces the Przanowski equation upon identifying
\be
z^1 = \txi\, ,
\qquad
z^2 = \alpha\, ,
\qquad
h=-2 K\, .
\ee
Having imposed \eqref{przxi0} and setting $z^3=\xi$, we can now solve for the remaining
components in terms of derivatives of $h$ in $z^1,z^2$ and of $K_3$ and $K_{\bar 3}$:
\bea
\label{K3123}
K_{3 \bar1} &=& \frac{h_{2\bar1}}{2 h_{\bar1}} -\frac{h_{1\bar1}}{h_{\bar 1}}\, K_{\bar 3} \, ,
\\
K_{3 \bar2}&=&
   \frac{h_{2\bar2}+2{\rm e}^{h}} {2 h_{\bar1}}
   +\frac{   h_{1\bar2}}{h_{\bar1}}\, K_{\bar3} \, ,
\\
   K_{3 \bar3}&=&   - \frac{1}{ |h_1|^2}\(\hf\,h_{2\bar2}+{\rm e}^{h}
+ K_{3 } h_{2\bar 1}+K_{\bar3}h_{1\bar 2}+2h_{1\bar 1}|K_{3 }|^2\)\, ,
\\
K^{ 3\bar 1}&=&-\frac{2 {\rm e}^{-h} }{h_{\bar 1}}\(h_{ 2\bar 1}+
(2h_{ 1\bar 1}-|h_{ 1}|^2)K_{\bar 3}\)  \, ,
\\
 K^{ 1\bar 1}&=& -\frac{ 4 {\rm e}^{-h}}{|h_{ 1}|^2}
\(\hf\, h_{ 2\bar 2}+{\rm e}^{h}
+K_{ 3 } h_{ 2\bar 1}+K_{\bar 3} h_{ 1\bar 2}+(2h_{ 1\bar 1}-|h_{ 1}|^2)|K_{ 3 }|^2\)\, .
\eea
The functions $K_3$ and $K_{\bar 3}$ are undetermined at this stage, however additional
conditions follow from requiring that $K_{ 3 \bar 3}$, $K_{ 3 \bar 1}$, $K_{ 3 \bar 2}$
are derivatives of $K$.
In particular, the phase $K_3/K_{\bar 3}$ can be determined as follows.
Since $h$ is related to the \kahler potential as in \eqref{hW} with $W=0$,
the variation of $W$ around zero should produce an eigenmode of the linearized
master equation \eqref{eqdPrz}. Therefore,
 \be
 \label{delhK3}
 \dPrz_{h}\left[ -2 \delta W_1(z^1) K_3 - \delta W(z^1) h_2 \right] = 0
 \ee
 for any holomorphic function $W(z^1)$.  Requiring the vanishing of the term proportional to $W''(z^1)$ leads to
 \be
 {\rm e}^{h} h_{\bar 1} K_3 - (2 {\rm e}^h + h_{2\bar 2}) K_{3\bar 1} + h_{2\bar 1} K_{3\bar 2} = 0
 \ee
which reduces, using \eqref{K3123}, to
\be
\frac{K_3}{K_{\bar 3}} = \frac{h_1}{\bar h_1}\, .
\ee
The vanishing of the term proportional to $W'$ provides an additional constraint,
but does not seem to allow to solve for the modulus of $K_3$ algebraically.

\section{Reconstructing the twistor lines from Przanowski's function}
\label{ap_sol}

In the presence of two commuting isometries, and assuming for simplicity that the anomalous
dimensions vanish, the twistor lines on $\cZ$ can be found from the Przanowski function $h$
as follows. The Darboux coordinate $\xi$ is defined globally
by the $\cO(2)$-valued moment map for the type 2 isometry, Eq.  \eqref{xiAR}.
Using \eqref{ressol}, this  can be written as
\be
\xi=\frac{h_2}{h_1}+\frac{{\rm e}^{h/2}}{h_1}\( t^{-1}-t\).
\ee
The other Darboux coordinates can be searched for as Taylor series in $t$,
\be
\begin{split}
\txii{+} &= z^2+\sum_{n=1}^\infty t^n\txii{+}_n,
\qquad
\ai{+} = z^1+\sum_{n=1}^\infty t^n\ai{+}_n.
\end{split}
\label{Tayxi}
\ee
Plugging in these expansions into \eqref{c-pot} and using \eqref{ressol}, \eqref{rescph}
and \eqref{defcoor}, one may derive the following condition
\be
\begin{split}
& \de \(z^1+\sum_{n=1}^\infty t^n\ai{+}_n\)
+\frac{h_2+{\rm e}^{h/2}\(t^{-1}-t\)}{h_1}\, \de \(z^2+\sum_{n=1}^\infty t^n\txii{+}_n\)
\\
& =\frac{1}{h_1 t}\(\de t +{\rm e}^{h/2} \de z^2+\frac{t}{2}\(h_\alpha\de z^\alpha-h_{\balp}\de \bz^{\balp}\)
+t^2 e^{h/2}\de \bz^{\balp}\).
\end{split}
\ee
Extracting the terms proportional to $\de t$, one obtains a set of algebraic equations which can be solved for $\ai{+}_n$, giving
\be
\ai{+}_n=-\frac{h_2}{h_1}\,\txii{+}_n+\frac{{\rm e}^{h/2}}{nh_1}\((n-1)\txii{+}_{n-1}-(n+1)\txii{+}_{n+1}\).
\label{solcoefalpha}
\ee
The remaining terms then produce a sequence of differential equations for the coefficients $\txii{+}_n$
\bea
\de\(\txii{+}_{2}+z^2+\bz^2\)+\txii{+}_2\(\de h-2\de\log h_1\)+{\rm e}^{-h}\(\de h_2-h_2\de \log h_1 \)  &=& 0,
\label{eqtxi2}
\\
\de\(\txii{+}_{n+1}+\txii{+}_{n-1}\)
+\((n+1) \txii{+}_{n+1}-(n-1)\txii{+}_{n-1}\)\( \frac12 \,\de h-\de\log h_1\)  &&
\label{eqtxin}
\\
+n {\rm e}^{-h/2}\txii{+}_n\(\de h_2-h_2\de \log h_1\)     &=&0,\qquad  n\ge 2.
\nonumber
\eea
This recursive system allows in principle to compute all the coefficients
in \eqref{Tayxi} in terms of the Przanowski
function $h$. In particular the integrability condition
for the first equation reduces to the master equation.


\providecommand{\href}[2]{#2}\begingroup\raggedright\endgroup

\end{document}